\documentclass[a4paper,12pt]{iopart}

\pdfoutput=1

\usepackage{amsfonts}
\usepackage{amssymb}
\usepackage{amsthm}
\usepackage{bbold}
\usepackage{subfig}
\usepackage{color}
\usepackage{fullpage}
\usepackage[pdftex]{graphicx}
\DeclareGraphicsExtensions{.pdf,.png,.mps,.ps,.eps,.svg,.jpg}

\definecolor{Gray}{gray}{0.92}

\begin{document}

\title{Stochastic delocalization of finite populations } \author{Lukas Geyrhofer, Oskar Hallatschek} \address{Biophysics and Evolutionary Dynamics Group, Max-Planck-Institute for Dynamics and Self-Organization, 37077 G\"ottingen, Germany} \ead{lukas.geyrhofer@ds.mpg.de, oskar.hallatschek@ds.mpg.de} \begin{abstract} The localization of populations of replicating bacteria, viruses or autocatalytic chemicals arises in various contexts, such as ecology, evolution, medicine or chemistry.  Several deterministic mathematical models have been used to characterize the conditions under which localized states can form, and how they break down due to convective driving forces. It has been repeatedly found that populations remain localized unless the bias exceeds a critical threshold value, and that close to the transition the population is characterized by a diverging length scale. These results, however, have been obtained upon ignoring number fluctuations (``genetic drift''), which are inevitable given the discreteness of the replicating entities.  Here, we study the localization/delocalization of a finite population in the presence of genetic drift.  The population is modeled by a linear chain of subpopulations, or demes, which exchange migrants at a constant rate. Individuals in one particular deme, called ``oasis'', receive a growth rate benefit, and the total population is regulated to have constant size $N$. In this ecological setting, we find that any finite population delocalizes on sufficiently long time scales. Depending on parameters, however, populations may remain localized for a very long time.  The typical waiting time to delocalization increases exponentially with both population size and distance to the critical wind speed of the deterministic approximation. We augment these simulation results by a mathematical analysis that treats the reproduction and migration of individuals as branching random walks subject to global constraints. For a particular constraint, different from a fixed population size constraint, this model yields a solvable first moment equation.  We find that this solvable model approximates very well the fixed population size model for large populations, but starts to deviate as population sizes are small. Nevertheless, the qualitative behavior of the fixed population size model is properly reproduced. In particular, the analytical approach allows us to map out a phase diagram of an order parameter, characterizing the degree of localization, as a function of the two driving parameters, inverse population size and wind speed. Our results may be used to extend the analysis of delocalization transitions to different settings, such as the viral quasi-species scenario.  \end{abstract}

\maketitle
\tableofcontents

\section{Introduction}

\begin{figure}[!t]
\begin{center}
\subfloat{\label{fig:realexamples:htv}\includegraphics[width=7.8cm]{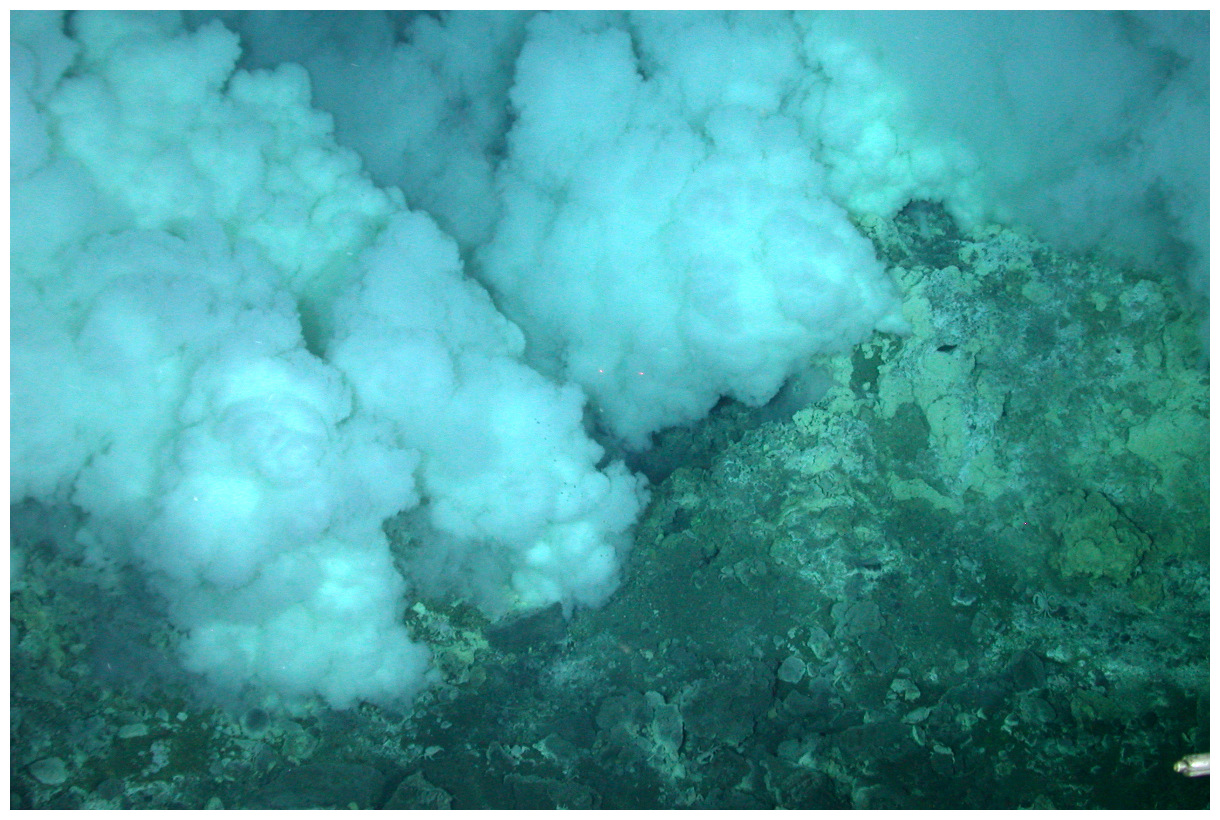}}
\subfloat{\label{fig:realexamples:oasis}\includegraphics[width=7.8cm]{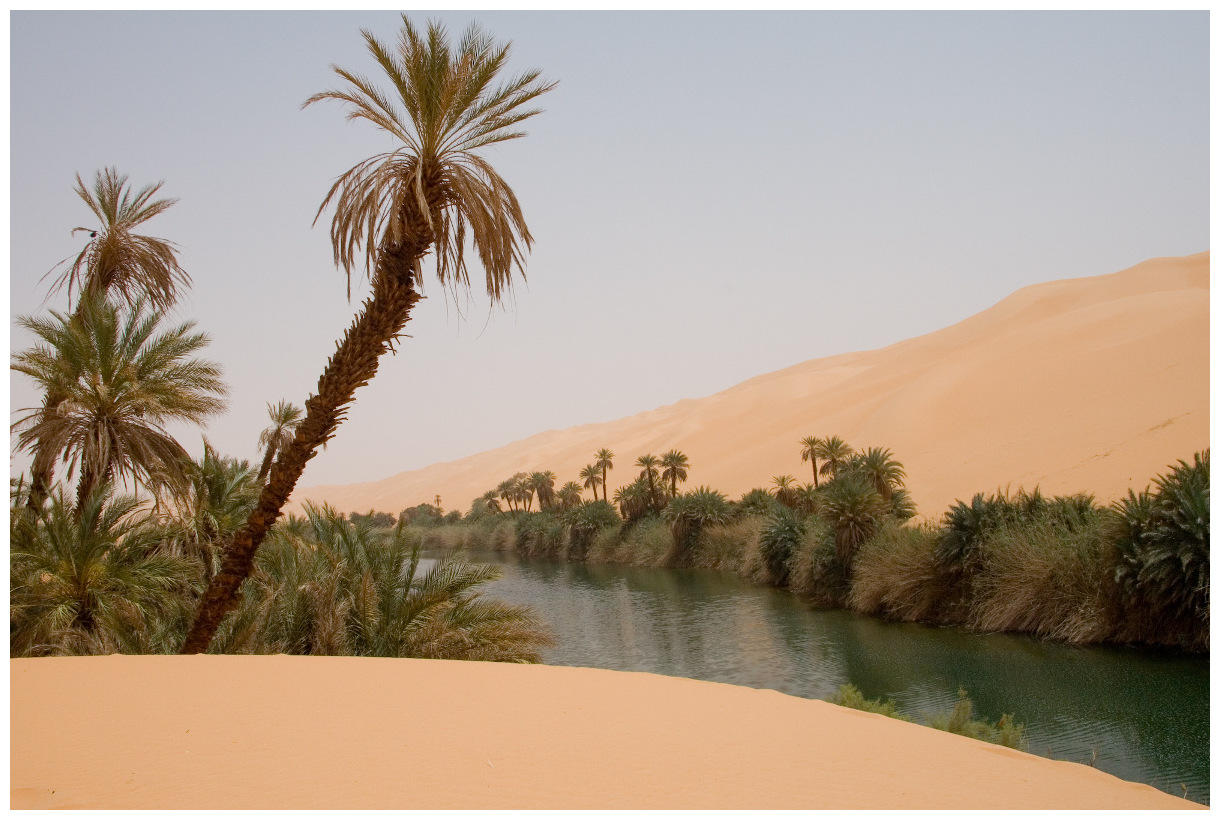}}
\caption{\label{fig:realexamples}\textbf{Examples for strongly heterogeneous environments.} (a) Chemo-synthetic bacteria near hydrothermal vents. (Photo: NOAA) (b) Localized growth in a desert: an oasis. (Photo: Luca Galuzzi)}
\end{center}
\end{figure}

Many population biological models simplify ecological complexity by assuming that habitats are homogeneous. Within these models, populations have a strong tendency to become evenly distributed in space as time proceeds. However, many biologically plausible habitats are strongly heterogeneous and provide gradients or localized spots of growth rather than homogeneous growth conditions.
For instance, hydrothermal vents are highly localized environments that may have played a key role in the evolution of life itself \cite{waechtershaeuser1990evolution}. Today chemo-synthetic bacteria using the heat, methane and sulfur of such vents form the basis of deep sea food chains, ranging from bacteria up to higher organisms as fish or octopuses \cite{tunnicliffe1991biology,tunnicliffe1998biogeographical} (Figure \ref{fig:realexamples:htv}).
An oasis in the desert is another example, where member species shape the environment for one another, thereby feeding back on the stability of the oasis (Figure \ref{fig:realexamples:oasis}).  When such heterogeneities are included in the population dynamics, the question arises whether populations will also become heterogeneously distributed, or localized, despite the mixing forces of diffusion or convection.

There have been several theoretical accounts to model the population dynamics at localized growth spots~\cite{dahmen2000life,shnerb2001extinction,desai2005quasispecies}. These studies focused in particular on the effect of a convection term, or wind, on the population localized at a growth spot, called the ``oasis''.
It has been found found that populations are localized as long as the convection forces remains below a certain \emph{critical} value. As the convection speed is increased, the population becomes more widely spread around the oasis. The spread of the population is characterized by a correlation length scale that diverges at the critical convection velocity, signaling a continuous phase transition to extended states.
This delocalization transition has also been investigated experimentally taking advantage of the fast growth rate of microbes~\cite{neicu2000extinction,lin2004localization}. In these experiments, bacterial colonies were grown under hazardous UV light. Only a small localized (and moving) plate shielded the colonies, thus creating a preferred growth environment similar to the oasis considered in the theoretical models.

Closely related localization phenomena have been studied in evolutionary biology. In this context, populations evolve in a heterogeneous fitness landscape, rather than a geological landscape. Deleterious mutations act like a convection force that tends to push populations away from the optimal (i.e. most fit) genotype. For small mutation rates, natural selection is able to maintain a population cluster localized at the fitness optimum.
For large mutation rates, however, populations drift away from the peak. The associated delocalization transition is called genetic meltdown in the population genetic literature \cite{lynch1993mutational,whitlock2003compensating}, respectively error threshold in the viral quasi-species literature \cite{eigen1971selforganization,eigen1989molecular,hermisson2002mutation,wilke2005quasispecies,wolff2009robustness}. 

\begin{figure}[tb]
\begin{center}
\includegraphics[width=10cm]{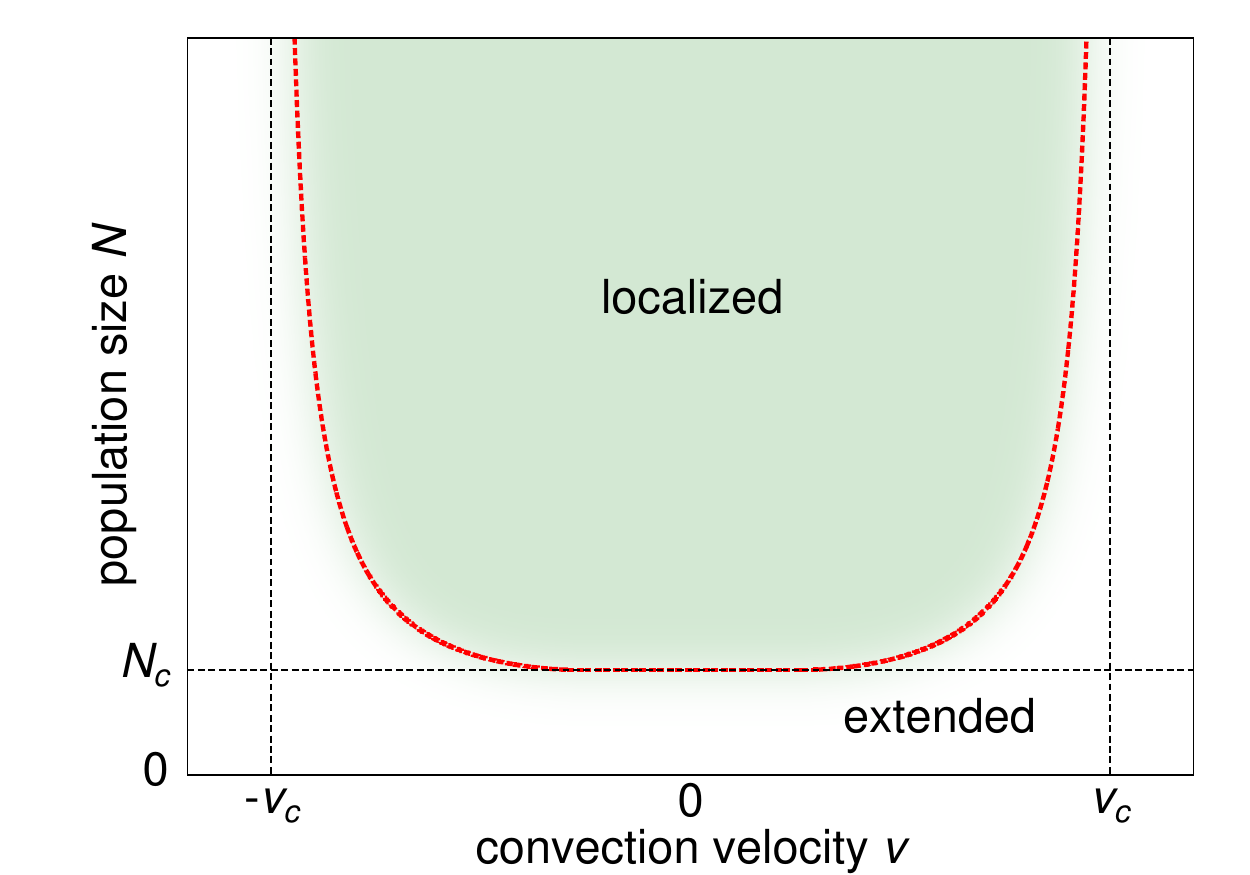}
\caption{\label{fig:locstates}
\textbf{Hypothesized phase diagram for finite populations localized at an ``oasis''.}
Deterministic calculations of a population localized at a favorable growth spot (``oasis'') predict a sharp delocalization transition: When the convection speed exceeds a critical value $v_c$, populations escape from the oasis.
Our stochastic simulations of populations of fixed size suggest that this picture has to be modified to account for number fluctuations. i) Populations always delocalize on sufficiently long time scales, due to number fluctuations.
The \emph{escape time}, however, depends exponentially on population size. Thus, delocalization is effectively unobservable in sufficiently large populations. ii) In the presence of convection, the escape times are strongly reduced. In an exactly solvable model with fluctuations but finite populations, a sharp transition line can be derived (red dashed line).
This line also describes qualitatively the crossover from small to very large escape times (compared to the coalescence time) in fixed population size models.}
\end{center}
\end{figure}

The results on both the ecological delocalization transition and the error threshold, have been obtained largely by a linear stability analysis of the localized state. This approach ignores number fluctuations, which are however inevitable in populations of discrete individuals. The fluctuations are generated by variations in the offspring number, which itself depend on factors such as environment, lifespan or genetic background.
The question thus arises how the standard view of the localization transition has to be modified to account for number fluctuations. We hypothesize that localized populations can be destabilized not only by convection but also by number fluctuations. Accordingly, we expect a modified phase diagram that depends on convection speed and population size, such as sketched in the tentative phase diagram (Figure~\ref{fig:locstates}).

To test our hypothesis, we study quantitatively the localization transition in the presence of number fluctuations. We first simulate a one-dimensional population localized near a single lattice site with enhanced growth rate. This oasis is surrounded by a homogeneous habitat of lower growth rate.
In order to fix the total population size, the excess population produced in each generation is removed by a (space-independent) death rate, which is adjusted in each time step. Our main goal is to reveal the impact of number fluctuations on the degree of localization, and the population distribution at steady state.
We augment our simulation results by a mathematical analysis based on branching random walks subject to a tuned global constraint. This solvable model has been shown to be a good approximation to fixed population size models as far as the description of noisy traveling waves is concerned \cite{hallatschek2011noisy}.

\section{Simulations}

\subsection{Simulation model}
\label{sec:simulation-model}

\begin{figure}[!t]
\begin{center}
\subfloat{\label{fig:oasismodel:substep1}\includegraphics[width=7.8cm]{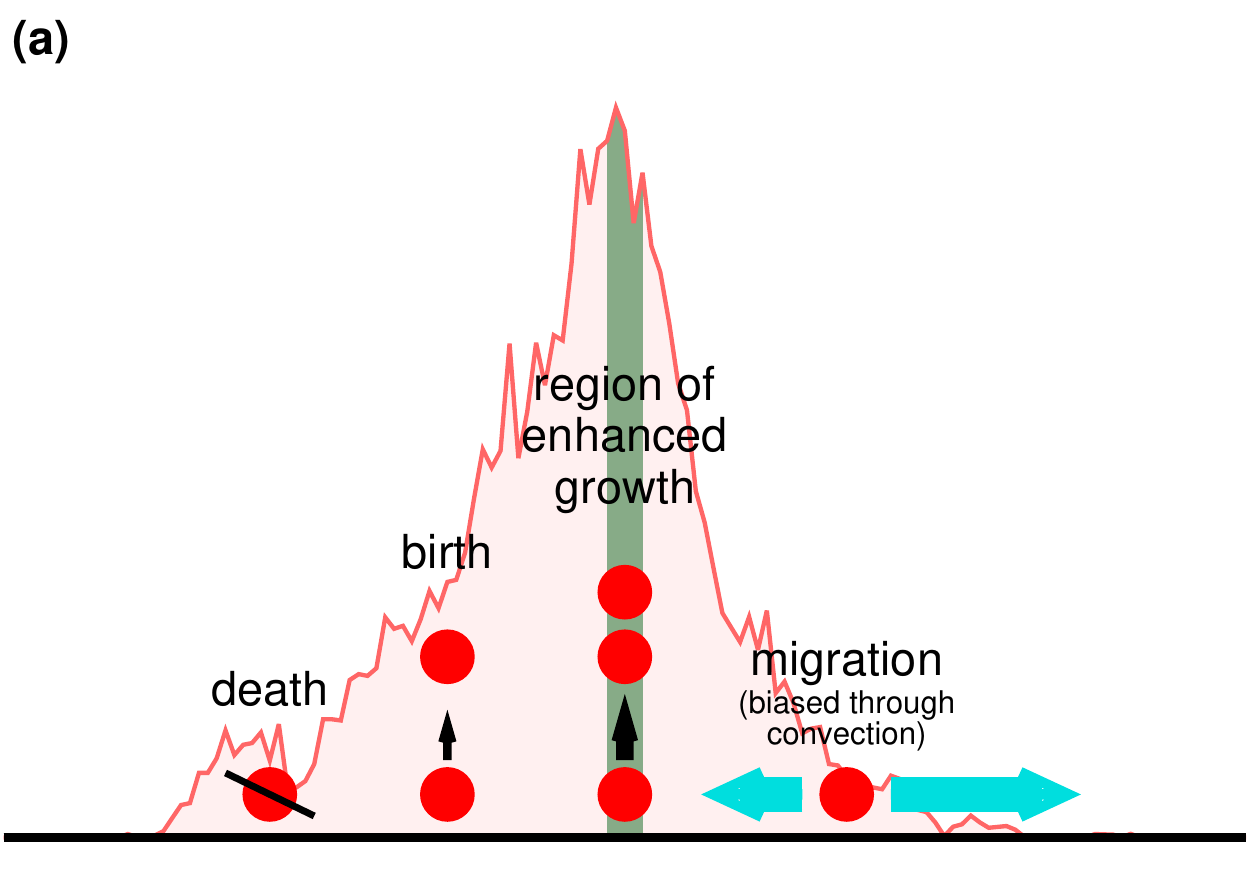}}
\subfloat{\label{fig:oasismodel:substep2}\includegraphics[width=7.8cm]{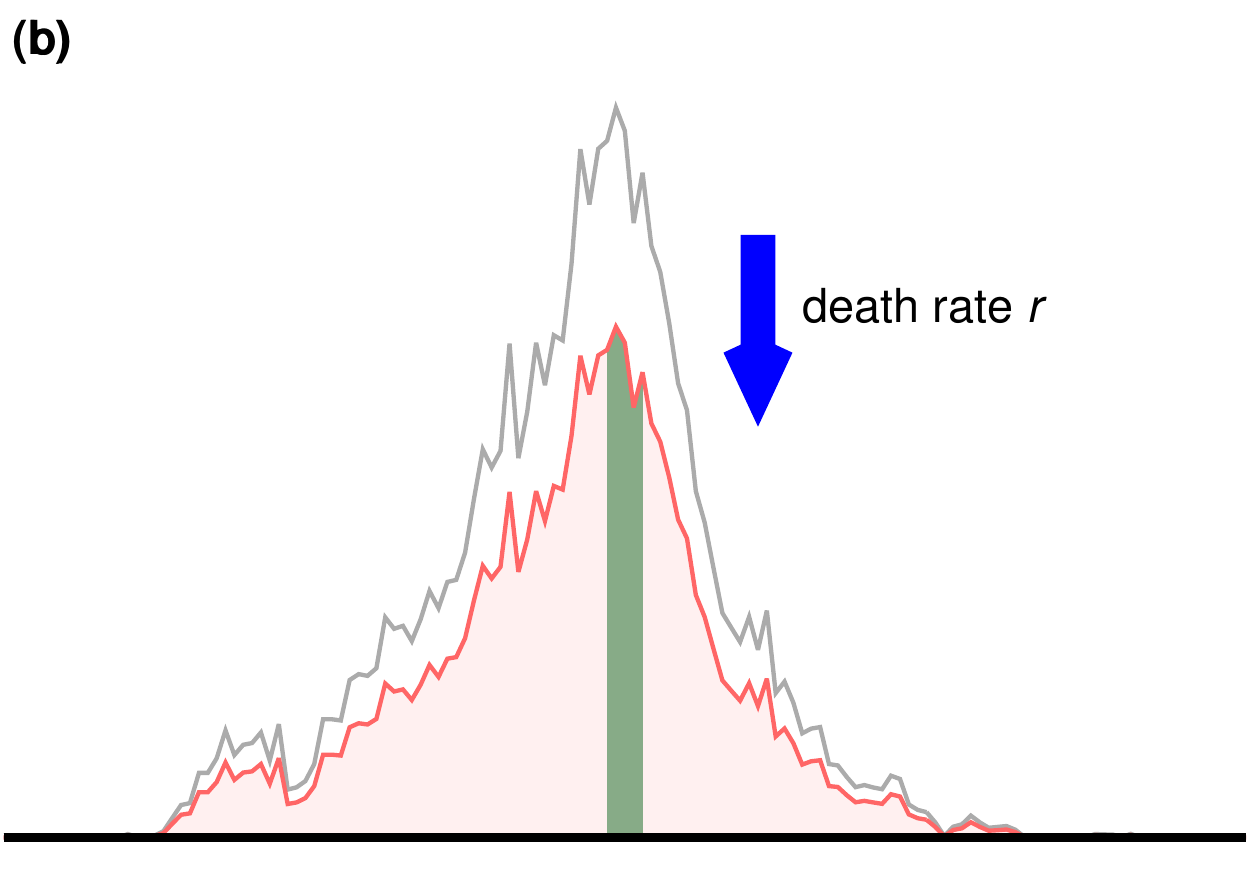}}
\caption{\label{fig:oasismodel}
\textbf{Stochastic dynamics of the ``oasis'' model.}
A single time-step in our simulations consists of two sub-steps. (a) Individuals migrate between neighboring lattice sites, or demes, in general with a bias in one direction to account for convection.
Number fluctuations are incorporated to account for the population turn over from one generation to the next. The subpopulation located at the oasis enjoys an increased reproduction rate $a$. This first sub-step effectively describes a branching random walk with a branching rate that depends on the location.
(b) In the second sub-step, the (global) population is regulated to comply with the constraint of fixed population size (see main text for details). }
\end{center}
\end{figure}

In our simulations, the population of reproducing individuals is distributed along a linear chain of sub-populations, called demes~\cite{kimura1964stepping,weiss1965mathematical}. These demes are considered to have a distance $d=1$ to their neighbors. The number of individuals in deme $i$ is denoted by $c_i$.
Each time step of duration $\epsilon$ (typically $\epsilon=0.2$) generations consists of the several sub-steps illustrated in Figure \ref{fig:oasismodel}. First, we add to each occupancy variable $c_i$ the deterministic change $\Delta^{\mbox{det}} c_i$ given by
\begin{equation}
  \label{eq:deterministic changes}
  \Delta^{\mbox{det}} c_i/\epsilon=\left(c_{i+1}+c_{i-1}-2c_i\right) + v \left(c_{i+1}-c_{i-1}\right)/2 +a \delta_{i0} c_0 \;.
\end{equation}
The first two terms on the right hand side account for biased migration: Individuals jump at rate $1+v/2$ from $i$ to $i+1$ and at rate $1-v/2$ from $i$ to $i-1$.  The last term represents deterministic growth at the oasis ($i=0$) at rate $a$.

Next, we account for a stochastic change by adding $\Delta^{stoch} c_i$ given by
\begin{equation}
  \label{eq:stochastic change}
  \Delta^{\mbox{stoch}} c_i/\sqrt{\epsilon}=\sqrt{2}\left(\mbox{Poisson}(c_i)-c_i\right) \;.
\end{equation}
Here, $\mbox{Poisson}(c_i)$ is a random number drawn from a Poisson distribution with mean $c_i$. The particular form of the noise in (\ref{eq:stochastic change}) ensures that i) occupancies do not drop below zero and ii) the generated variance in occupancy is given by $2\epsilon c_i$.
The variance produced thus equals the expected variance in a time $\epsilon$ for a (large) population of individuals that duplicate and die at the rate $1$ (in our computational unit of time). We note that the natural alternative of directly simulating individuals that duplicate and die is computational much more expensive if $\epsilon c_i$ is much larger than $1$. 

After the previous sub-steps, the population size will typically be larger than the pre-set value $N$. In order to restore the population size constraint, we add to each variable $c_i$ the change 
\begin{equation}
  \label{eq:culling}
  \Delta^{\mbox{cull}}c_i/\epsilon=-r c_i 
\end{equation}
with a (global) death rate
\begin{equation}
  \label{eq:r}
  r=1-N/\sum_i c_i \;.
\end{equation}
The rescaling of the occupancies by the factor of $1-r$ effectively represents a homogeneous culling of the population that is independent of the location. Notice that, although our simulations account for the fluctuations induced by the discreteness of the individuals, the occupancies $c_i$ are generally non-integer. The approximations involved are equivalent to the diffusion approach widely used in population genetics~\cite{kimura1985neutral,drossel2001biological}: As long as populations are large ($N\gg1$) and the temporal changes exhibit the same mean and variance as the discrete particle system, the deviations between both approaches should be small in the limit $\epsilon\to 0$. This means in particular that the results become insensitive to the precise noise distribution used, as long as the first two moments are correct. 

We impose periodic boundary conditions, as if demes were arranged on a circle like in the experiments of Ref.~\cite{neicu2000extinction}. The number of demes is chosen large enough to ensure that population fits the simulation box. Time averages are taken after the simulations have become independent of their localized initial conditions.

\subsection{Simulation results}

\begin{figure}[!t]
\begin{center}
\subfloat{\label{fig:snapshots:nowind}\includegraphics[width=7.8cm]{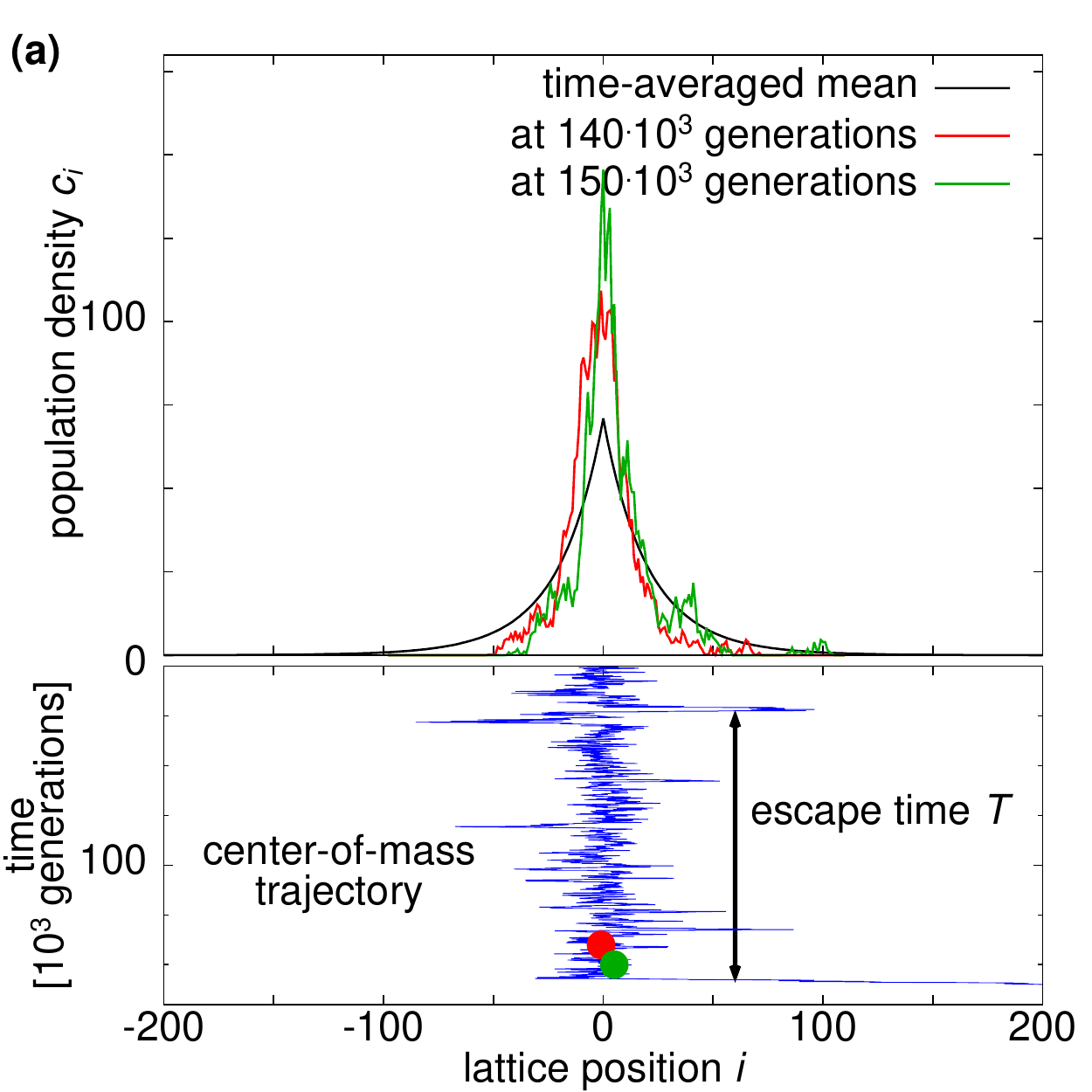}}
\subfloat{\label{fig:snapshots:wind}\includegraphics[width=7.8cm]{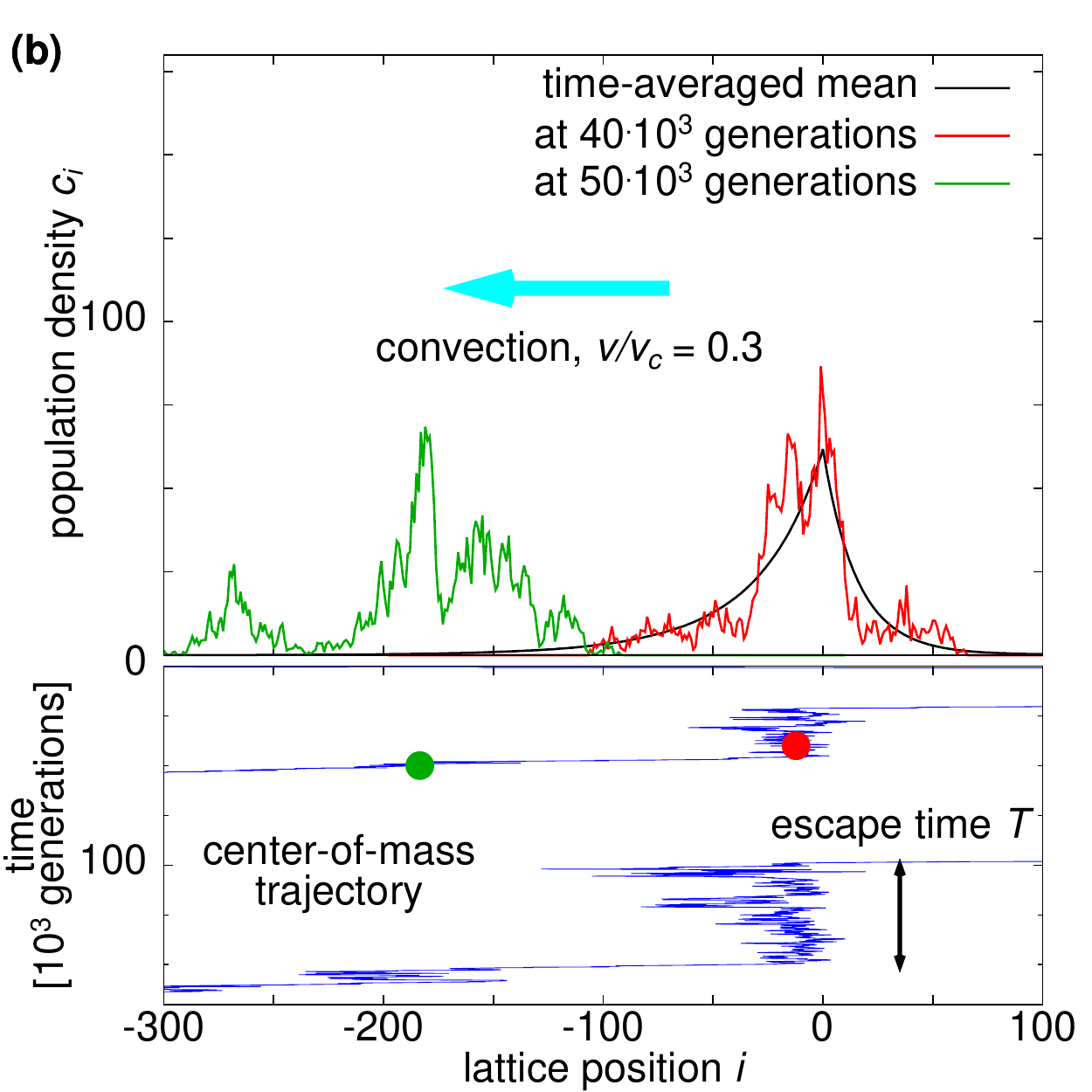}}
\caption{\label{fig:snapshots}
\textbf{Snapshots of simulations with fixed population size.}
Figure (a) depicts snapshots for our simulation model, in which  individuals carry out an unbiased branching random walk subject to the global constraint of fixed population size (see Figure~\ref{fig:oasismodel} for a description of the time step).
All lattice sites (labeled by an index $i$) are equivalent, except for deme $i=0$ where the growth rate is increased by $a=0.1$. The resulting density profiles are typically quite narrowly centered around the oasis (c.f. green and red lines). The average profile (black line) is much broader than typical profiles, due to large excursions of the center-of-mass.
These excursions can be seen in the center-of-mass trajectory at the bottom part of the figure. Notice that the population occasionally escapes from the oasis, due to a rare large fluctuation. Such delocalizations are recorded when the occupancy at the the oasis drops below a critical value, $c_0<1$. In the convection-less case, the population returns quickly after a random excursion (e.g. the first delocalization event in (a)). With convection, (b), the population usually traverses the (periodic) simulation box after a delocalization event until it reaches the oasis again. The escape time between two delocalizations is indicated by a black arrow.
The green and red bullets indicate the spatio-temporal position of the center-of-mass corresponding to the red and green snapshots in the upper part of the figure. The convection speed in (b) is set to $30\%$ of the critical convection speed $v_c=ad$, at which delocalization is observed in the deterministic limit~\cite{dahmen2000life}.
Notice that convection leads to more frequent escapes from the oasis. Simulation parameters were $a=0.1$, $N=3000$, $\epsilon=0.2$.}
\end{center}
\end{figure}

Snapshots of the population density are shown in Figure \ref{fig:snapshots}. For the chosen parameter values, fluctuations in the population density are clearly visible. The time traces of the center-of-mass trajectory (lower half of Figure \ref{fig:snapshots}) indicate that the population peak fluctuates wildly around the oasis, until it eventually escapes.
The escape time (or delocalization time) is much greater in the case without convection (a) than with convection (b), but it eventually occurs. In fact, no sharp transition between localized and delocalized populations is detectable in our noisy system. If the population size is large enough, the delocalization is however unobservable within reasonable (simulation-) time. For definiteness, we say the population is on the oasis, when at least one individual is present there, $c_0>1$, and off the oasis, when this condition is not met.
We checked several alternative conditions, they yield similar or comparable results (see appendix \ref{apdx:condition}). Time averaged quantities (i.e. occupancies) are then obtained by averaging only over states with $c_0>1$. 

We measured in detail the time-averaged death rate $\rho\equiv\overline{r}/\left(a^2d^{2}/4D\right)$, scaled by its value $a^2d^{2}/4D$ in the deterministic no-convection limit. If the overall death rate is large, a population can only be sustained in the vicinity of the oasis where net-growth is positive. Consequently, populations will be strongly localized at the oasis for $\rho=O(1)$.  For smaller and smaller death rates, the population will be able to explore larger and larger parts of the habitat. When the death rate eventually changes sign (thus becoming a growth rate), growth becomes possible throughout the entire habitat. As a consequence, the population distribution becomes necessarily extended in space. Therefore, the death rate $\rho$ may be considered as an order parameter that measures the degree of localization of the population at the oasis. Consistent with this interpretation, the mean field theory predicts that $\rho$ continually decays from $1$ to $0$ as the convection speed is increased from $0$ to the critical value $v_c=ad$.

\begin{figure}[!tb]
\begin{center}
\includegraphics[width=10cm]{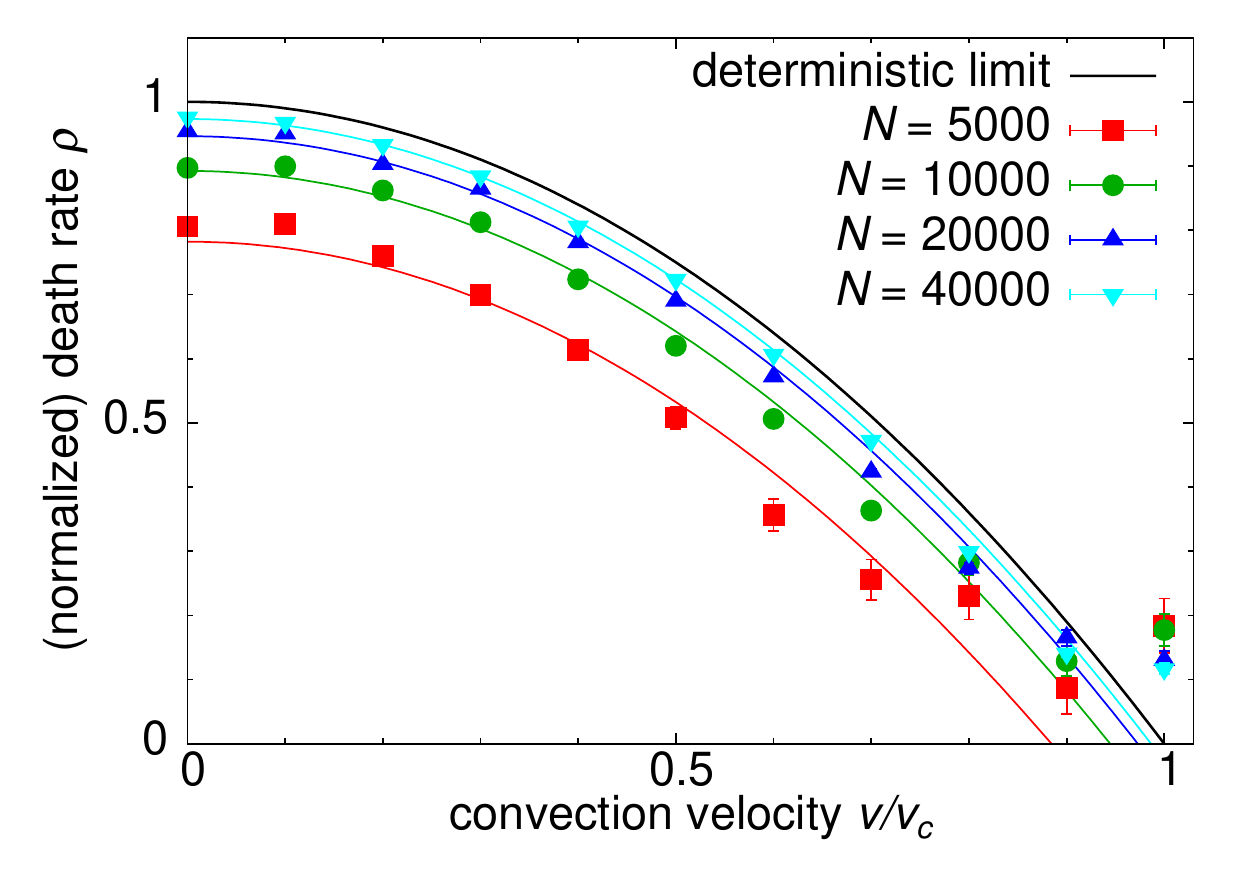}
\caption{\label{fig:deathrates}
\textbf{Mean death rates for varying population size and convection speed.}
In our simulations, the death rate $r$ is the factor with which the population has to be scaled down to keep its global size constant (c.f. section \ref{sec:simulation-model}). Here, we display the scaled death rate  $\rho = \overline{r}/\left(a^2d^2/4D\right)$ normalized by its deterministic no-convection limit $a^2d^2/4D$.
According to the deterministic limit, the death rate should decay with increasing convection speed according to $\rho=1-(v/v_c)^2$ (solid black line) up to the critical convection speed $v_c=ad$~\cite{dahmen2000life}. For finite populations the death rates are shifted to lower values roughly by the constant $2/N$, as predicted by the analytical approach explained in section  \ref{sec:theoreticalapproach}.
As a consequence, smaller convection speeds are required for the death rates to cross $0$, which signals the delocalization of the population. Our time average only includes states in which there is at least one individual at the oasis, $c_0>1$. For convection speeds larger or equal than the deterministic critical velocity, this is just a tiny minority of all states.
Hence, the points for large convection speeds (close to $v_c$) are slightly biased.
Solid lines are obtained by our theoretical approach explained in section \ref{sec:theoreticalapproach}, in particular by the relation (\ref{eq:sva-r}).
}
\end{center}
\end{figure}

The death rates $\rho$ obtained from simulations are depicted in Figure~\ref{fig:deathrates}, across a range of population sizes and convection velocities bounded by $v_c=ad$ for which delocalization is observed in the deterministic limit. As expected from the mean field theory, we find that the death rate continually decays to zero with increasing convection $v$. In addition,  we find that death rates decrease when we decrease the population size.
For any finite population, death rates go to $0$ at convection speeds significantly smaller than expected from the mean field predictions. Thus, increasing the demographic noise tends to destabilize the localized populations.

\begin{figure}[!tb]
\begin{center}
\subfloat{\label{fig:waitingtime:histo}\includegraphics[width=7.8cm]{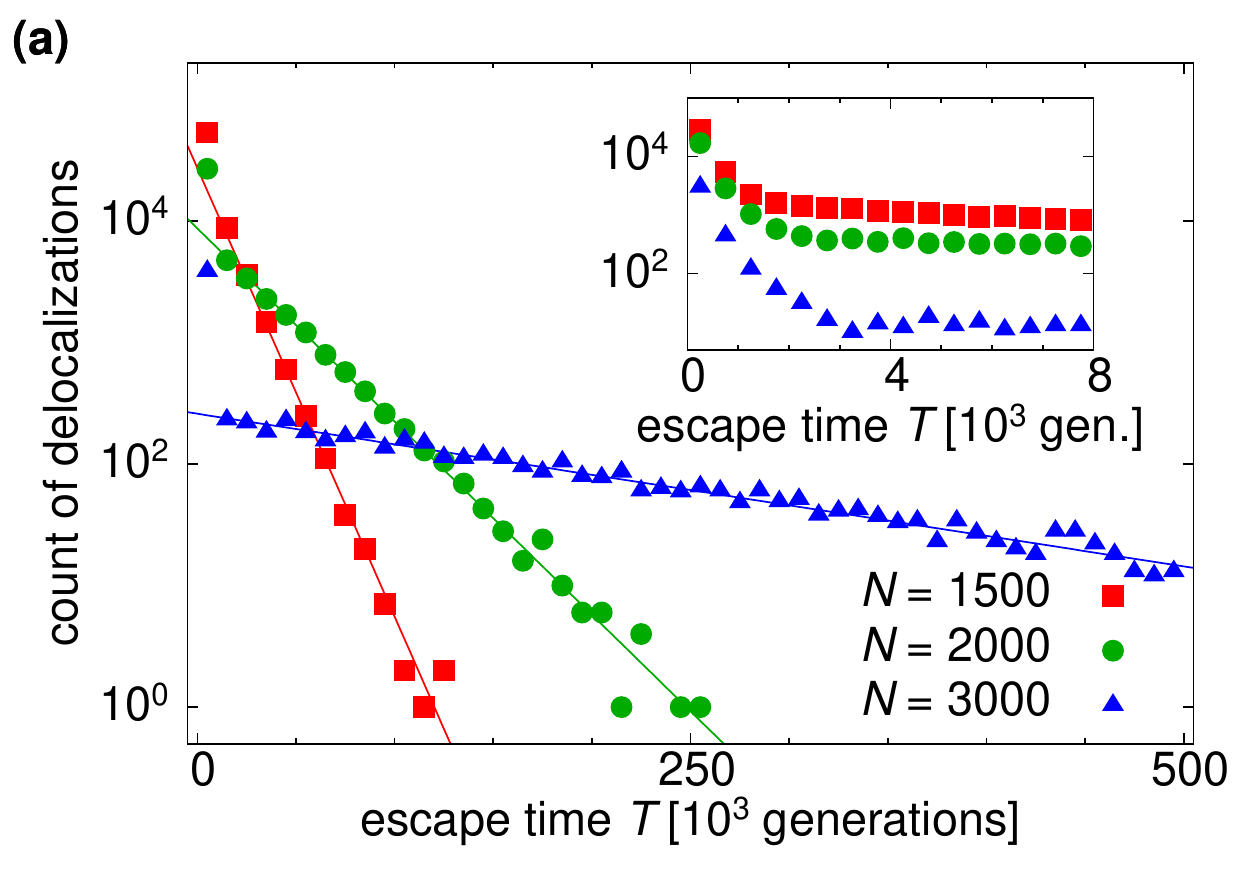}}
\subfloat{\label{fig:waitingtime:chartime}\includegraphics[width=7.8cm]{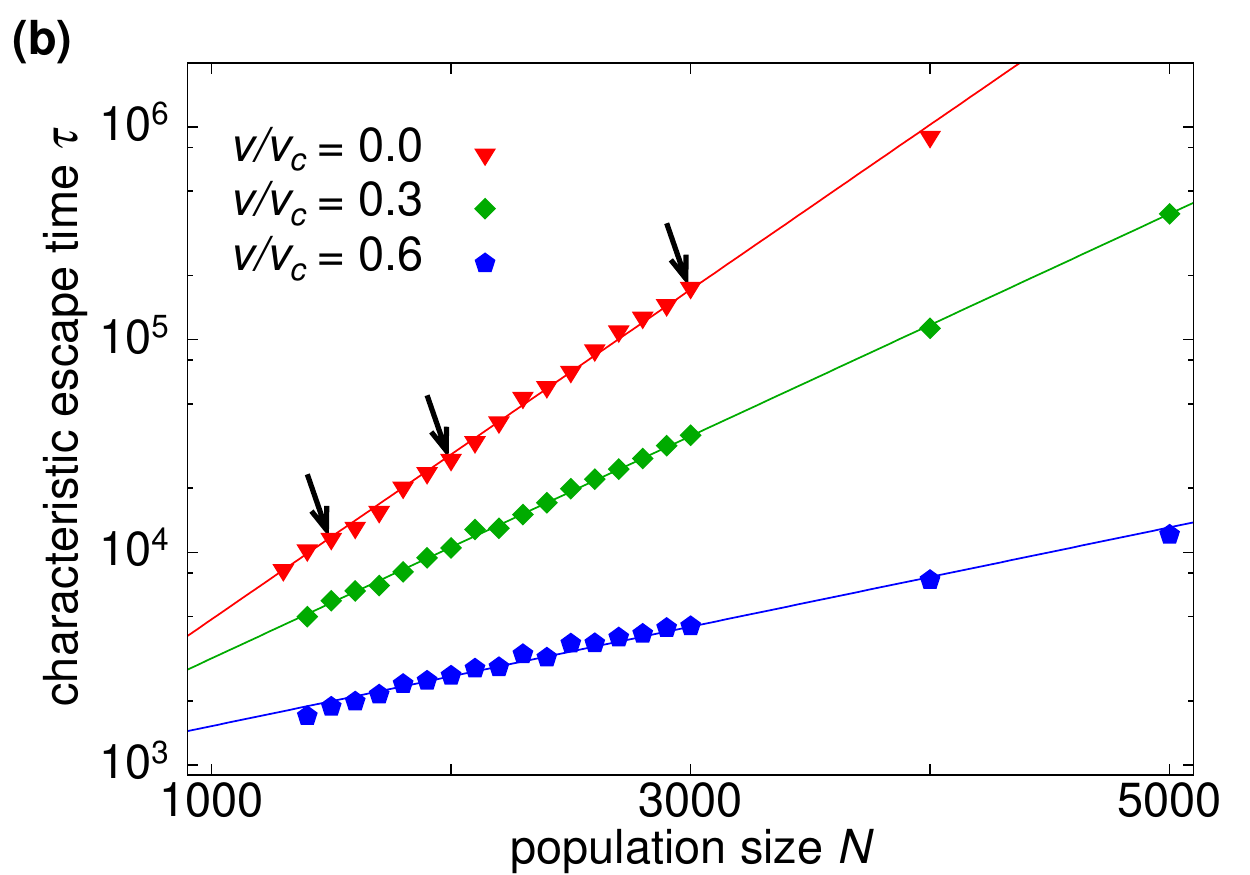}}
\caption{\label{fig:waitingtime}
\textbf{Distribution of escape times $T$ measured in simulations of fixed population sizes.}
(a) Distribution of escape times $T$ for different population sizes in the absence of convection. In all cases, the distribution is close to exponential. Deviations at early times (inset) are attributed to shape fluctuations, see main text.
From the exponential tail of the distribution, a characteristic escape time $\tau$ can be extracted. This time strongly depends on the population size and convection velocity, as shown in figure (b). Solid lines in figure (b) represent the fit in (\ref{eq:chartime}). The three black arrows indicate the characteristic times $\tau$ of the distributions shown in (a). The simulation results were obtained using parameter values $a=0.1$, $\epsilon=0.2$.}
\end{center}
\end{figure}

Another more direct measure for the degree of localization is the escape time $T$ between a localization and the consecutive delocalization event, as indicated in the the center-of-mass trajectories in Figure \ref{fig:snapshots}. Histograms of escape times for various population sizes are shown in Figure \ref{fig:waitingtime:histo}. Notice that the escape time distribution is to a good approximation exponential,
\begin{equation}
\Pr(T)\sim \exp\left(-\frac{T}{\tau}\right) \;,
\label{eq:waitingtimedistr}
\end{equation}
indicating a memory-less delocalization process with a characteristic escape time $\tau$. 
Deviations from the exponential distribution are discernible only on short times (inset in Figure \ref{fig:waitingtime:histo}). These deviations result from fluctuations in the density profile that temporarily satisfy the delocalization condition $c_0<1$. After a relaxation process, the occupancy at the oasis is restored. Genuine delocalization is therefore described by only the exponential part of the distribution.
The characteristic escape time $\tau$ can be extracted from the slope on a semi-logarithmic plot. This quantity $\tau$ is depicted in Figure~\ref{fig:waitingtime:chartime} as a function of population size for various convection velocities. Notice that $\tau$ decreases when either the population size is decreased (because the relative strength of fluctuations increases), or the convection speed is increased (because convection removes individuals from the oasis, thereby spreading the population).
Numerically, our observations can be summarized by (Figure \ref{fig:waitingtime:chartime})
\begin{equation}
\tau \sim \exp\left(\frac{N}{\eta}\right) 
\label{eq:chartime}
\end{equation}
with $\eta\approx \beta\left(1-v/v_c\right)^{-1}$ and $\beta\approx560$.

In summary, our numerical simulations indicate that, due to number fluctuations, populations inevitably escape the ``pinning'' at the oasis on sufficiently long time scales. The typical escape times, however, increase exponentially with population size, and thus delocalization is effectively unobservable for sufficiently large populations in the absence of convection. Our goal for the following is to develop an analytical framework to predict a critical population size in order to observe localization for a given convection velocity.
This will be used to support the phase diagram in Figure \ref{fig:locstates}.

\section{\label{sec:theoreticalapproach}Analytical approach}
While our simulations are discrete in both space and time, they should be amenable to a continuous description if the time step $\epsilon$ and the reaction rate $a$ are small enough. Under these assumptions, the stochastic dynamics can be approximated by a reaction-diffusion equation of the form
\begin{equation}
\partial_t c(x,t) = D\partial_x^2c(x,t) + v\partial_xc(x,t) + K(x,t)\;.
\label{eq:diffconvreact}
\end{equation}
Here, $D$ is a diffusion coefficient, which has to be set equal to $1$ deme$^2$ per generation to match our simulations.  The second term on the right hand side describes convection with velocity $v$. The last ``reaction'' term $K(x,t)$ summarizes all stochastic and deterministic changes due to the turn-over of the population. An example of the successful application of such reaction-diffusion equations in population biology is the Fisher-Kolmogorov wave equation, in which the reaction term describes logistic growth \cite{fisher1937wave,kolmogorov1937study}.
This equation exhibits traveling wave solutions, which have been used to describe range expansions, epidemic outbreaks or the spreading of beneficial mutations~ \cite{kenkre2003applicability,kenkre2004results,hallatschek2011noisy}.

\subsection{\label{sec:detlim}Deterministic limit}
In the limit of very large populations, we can use a deterministic reaction term to describe the population dynamics in the oasis model,
\begin{equation}
  \label{eq:deterministic-reaction}
  K(x,t)=\left[\alpha \delta(x)-r \right]c(x,t)\;.
\end{equation}
The first term approximates the growth rate benefit provided at the oasis by a Dirac-delta-function $\delta(x)$. The pre-factor $\alpha$ represents the product of the increased growth rate $a$ at the oasis and the linear size $d$ of the oasis, $\alpha=ad$. The constant $r>0$ represents the death rate, which is tuned such that (\ref{eq:diffconvreact}) admits a stable steady state.

The analysis of the resulting reaction-diffusion system has been carried out in Ref.~\cite{dahmen2000life}. It is found that the rescaled death rate $\rho\equiv 4D r/\alpha^2=1$ in the absence of convection, $v=0$, and continually decays to $0$ as the convection speed approaches the critical speed $v_c=\alpha$, 
\begin{equation}
  \label{eq:rho-deterministic}
  \rho(v)= 1-(v/v_c)^2\;.
\end{equation}
It can be seen from Figure~\ref{fig:deathrates} that this mean field result clearly overestimates the death rates measured in the stochastic simulations.

The localized steady state profile is found to be given by 
\begin{equation}
  \label{eq:ss-profile-deterministic}
  c_{st}(x)\sim\exp\left[(-\alpha|x|+v x)/(2D)\right]\;.
\end{equation}
The density profile in equation (\ref{eq:ss-profile-deterministic}) has characteristic width $\xi(v)$, given by
\begin{equation}
  \label{eq:correlation-scale}
  \xi(v)=\frac{2D}{\alpha -v} \;,  
\end{equation}
which diverges as a power law as the convection speed approaches the critical speed. This diverging length scale indicates that convection driven delocalization is a continuous phase transition in the deterministic limit.

\begin{figure}[!t]
\begin{center}
\subfloat{\includegraphics[width=7.8cm]{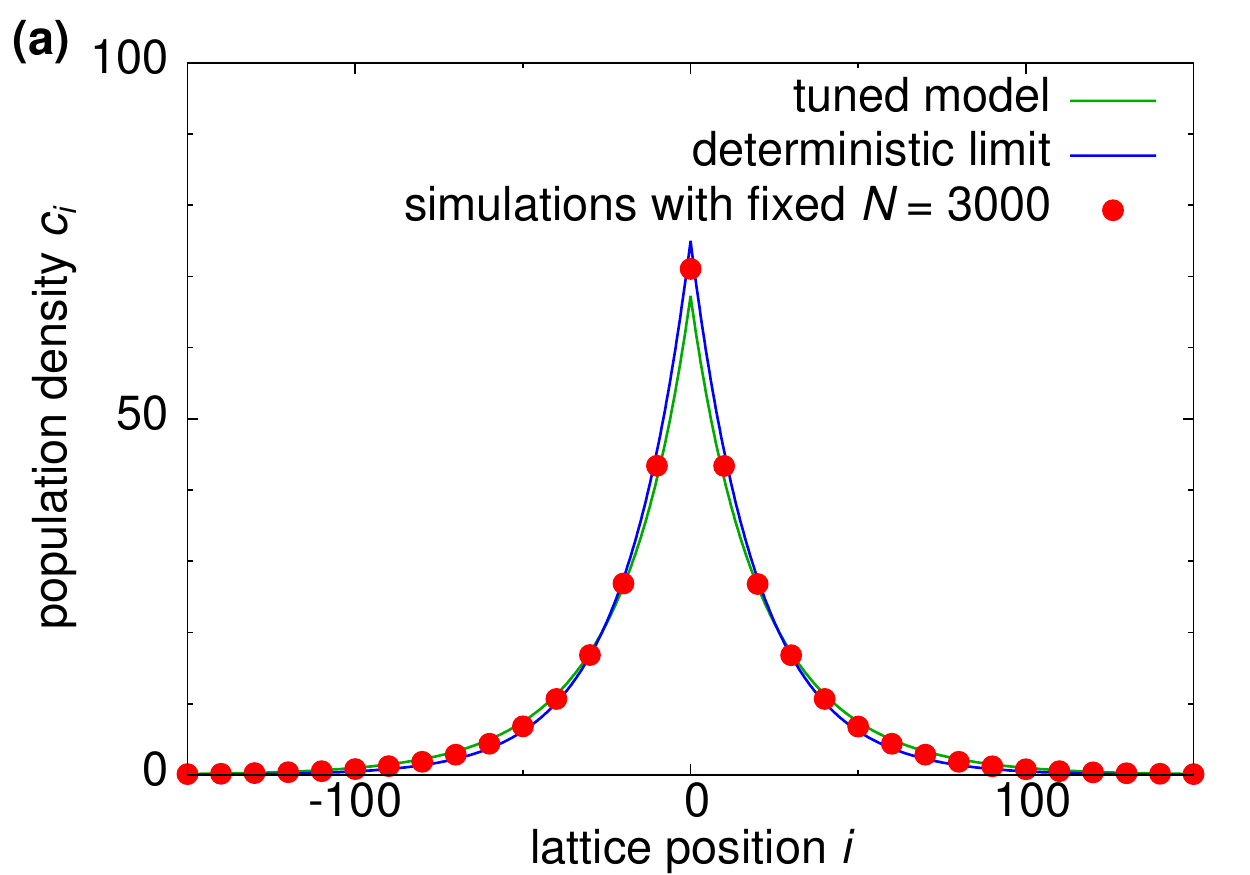}}
\subfloat{\label{fig:densshapes:convection}\includegraphics[width=7.8cm]{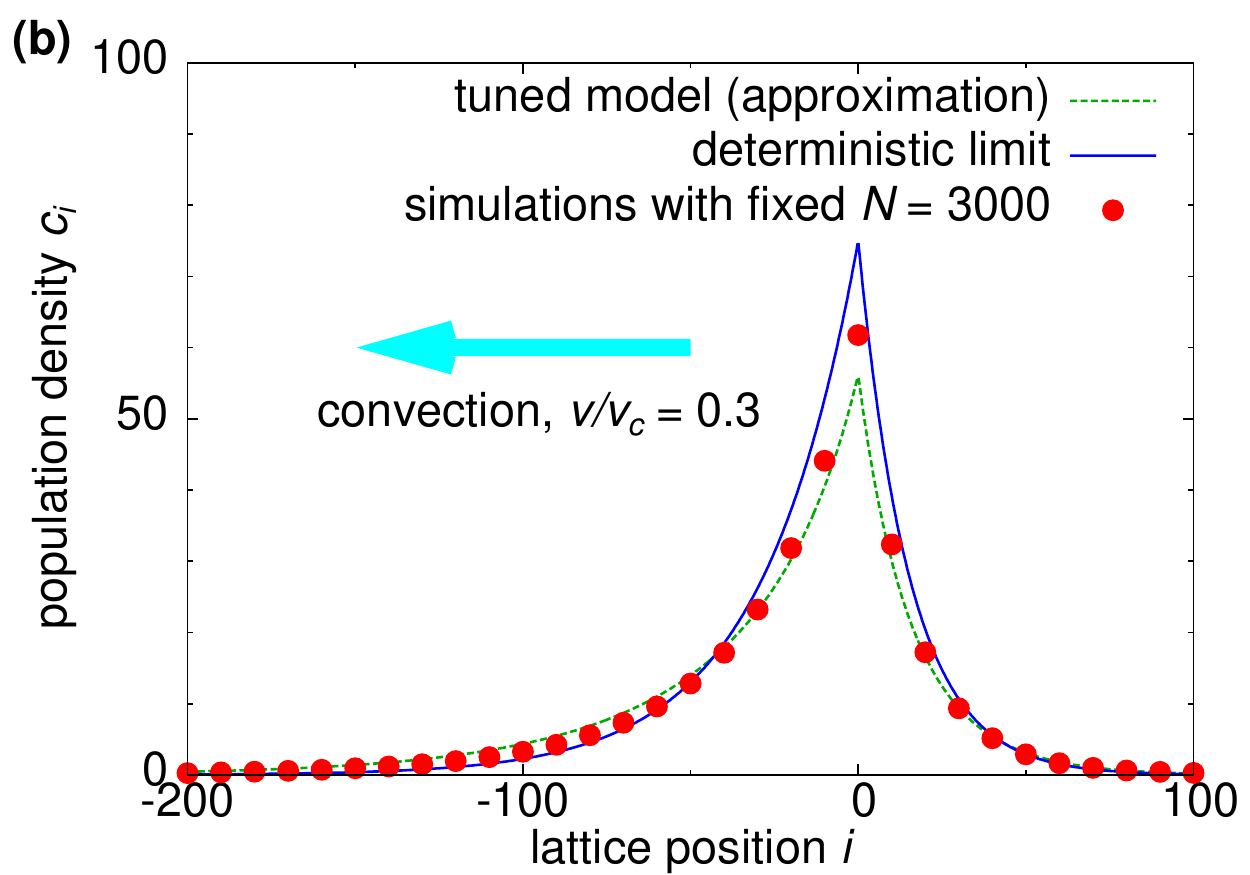}}
\caption{\label{fig:densshapes}
\textbf{Density profiles obtained for different models and approximations.}
(a) and (b) correspond to cases without and with convection, respectively. Symbols represent simulations with population size constraint. Green lines are predictions for the tuned constraint model, which we can derive analytically in the convection-less case (a).
When including convection, only approximations are possible (b) (c.f. section \ref{sec:withwind}). Blue lines represent the deterministic mean field approximations. For the parameters used ($a = 0.1$, $\epsilon= 0.2$, $N=3000$), all approaches yield quite comparable density profiles. Significant differences arise however in the death rates even at these large populations, c.f. Figure~\ref{fig:Nvsrho}.}
\end{center}
\end{figure}

The predicted density profiles are depicted in Figure \ref{fig:densshapes}, together with stochastic simulation data. The agreement is for the chosen parameter sets rather good, which is in contrast to the discrepancy between predicted and measured death rates.

\subsection{General approach to finite population sizes}
\label{sec:finite-popul-sizes-1}
Modeling the spatial dynamics of finite populations typically leads to stochastic non-linear reaction diffusion equations: A non-linearity is necessary to fix the population size, and stochasticity arises due to the finiteness of the population. As a consequence, these models are generally hard to treat analytically.
Systematic approximation schemes, such as perturbation expansions, often fail, as the example of noisy traveling waves demonstrates~\cite{vansaarlos2003front}. 

In a recent study, we have shown, however, that there is one particular form of the non-linearity that allows treating these and similar problems analytically~\cite{hallatschek2011noisy}. We now review our methodology, and describe how the same method can be applied to diverse situations, including both ecological and evolutionary problems.

Our method applies to a whole class of models of branching random walks subject to a global constraint. A computational time-step of size $\epsilon$ of such models consists of two sub-steps, which are very similar to the sub-steps of our simulation model. The first sub-step evolves the population density according to a branching random walk, which is regulated in the second sub-step to comply with a global constraint. Specifically, the first sub-step takes the concentration field $c(x,t)$ to an intermediate state 
\begin{equation} 
  \widetilde{c}(x,t+\epsilon)-c(x,t) = \epsilon\mathcal{L}c(x,t) + \sqrt{2\epsilon c(x,t)}\eta(x,t)\;.  \label{eq:oasismodel} 
\end{equation} 
Here, the first term on the right-hand side describes the deterministic changes, which are due to net-growth and net-migration in our oasis model. In a time step $\epsilon$, the associated change in density is given by $\epsilon \mathcal{L}$, where the Liouville operator $\mathcal{L}$ reads 
\begin{equation} 
  \mathcal{L} = D\partial_x^2+v\partial_x+\alpha\delta(x)- r \;.  \label{eq:linoperator} 
\end{equation} 
Whereas the first two (particle conserving) terms describe biased diffusion with diffusivity $D$ and average bias $v$, the last two reaction terms account for an overall negative growth rate $-r$ except for the enhanced growth at the oasis. We note that, formally, we can reinterpret our model in an evolutionary context, if we consider $x$ to be a coordinate measuring either a quantitative trait with continuous characteristics, such as height, skin, or body mass, or a co-ordinate measuring the genetic distance. In both cases, the biased diffusion corresponds to the random change due to mutations that often tend to favor one direction. The delta-function growth spot would model an optimal phenotype which enjoys an increased fitness. The delocalization threshold would have to be considered as Eigen's error threshold upon which selection is not able to maintain the optimal phenotype. More complicated quasi-species models can be easily devised, for instance, by considering a whole spectrum of mutations, recombination or more complex fitness landscape~\cite{good2012distribution,neher2011statistical,drossel2001biological}.

The stochastic second term on the left hand side in (\ref{eq:oasismodel}) accounts for the randomness in the reproduction process.  The random function $\eta(x,t)$ has vanishing mean and is uncorrelated across space and across the discrete time-steps, 
\begin{equation}
  \label{eq:noise}
  \overline{\eta(x,t)\eta(x',t')}=\delta(x-x')\delta_{(t/\epsilon),(t'/\epsilon)}\;.
\end{equation}
The amplitude $\sqrt{2\epsilon c(x,t)}$ of the noise is chosen such that the produced variance represents the number fluctuations generated in the time span $\epsilon$ by a population of size $c(x,t)$, in which individuals branch and die at the same rate $1$.
The amplitude of the noise is thus consistent with the source of stochasticity in our simulations described in section \ref{sec:simulation-model}.

In the second sub-step, 
\begin{equation}
c(x,t+\epsilon) = \widetilde{c}(x,t+\epsilon)(1-\lambda)\;,
\label{eq:globalconstraint}
\end{equation}
the population density $\widetilde{c}(x,t+\epsilon)$ is rescaled by a factor $1-\lambda$. Here, $\lambda$ has to be chosen in such a way that the density $c(x,t+\epsilon)$ at the end of the time step satisfies the constraint 
\begin{equation}
1 = \int dx~u(x)c(x,t+\epsilon) \;.
\label{eq:generalizedconstraint3}
\end{equation}
The weighting function $u(x)$ is at this point arbitrary. If one chooses $u(x)=1/N$ one obtains a fixed population size model, as we have simulated above (up to the lattice discreteness). For any other choice of $u(x)$ the population size is not fixed, but becomes a fluctuating quantity.

Notice that it is the constraint (\ref{eq:generalizedconstraint3}) that introduces a non-linearity in the problem, because $\lambda$ depends on $\widetilde c$. The resulting noisy nonlinear reaction diffusion problems are intractable for general weighting functions $u(x)$.
This can be seen from the first moment equation for the mean $\overline c(x,t)$, for which one can derive the equation of motion
\begin{equation}
\partial_t\overline{c} = \left(\mathcal{L}-2u\right)\overline{c} - \overline{\left\langle c\vert \left(\mathcal{L}^\dagger -2u\right)u\right\rangle c}\;.
\label{eq:firstmoment}
\end{equation}
A detailed derivation of (\ref{eq:firstmoment}) can be found in \cite{hallatschek2011noisy}. Notice that the second term on the right hand side in general involves the second moment, $\overline{c(x,t)c(y,t)}$. The equation of motion for the second moment, on the other hand, involves the third moment, and so on.
The resulting infinite hierarchy of moment equations is intractable unless one is willing to introduce an ad-hoc truncation scheme.

However, it is possible to obtain a closed first moment equation for a special choice of the weighting function. Suppose, the solution  $u_*(x)$ of the nonlinear differential equation
\begin{equation}
\left(\mathcal{L}^\dagger -2u_*\right)u_* \stackrel{!}{=} 0 
\label{eq:ustar}
\end{equation}
exists, and is taken as our choice of the weighting function. Then, the non-linear second term of the moment equation (\ref{eq:firstmoment}) vanishes identically. As a consequence, the dynamics of the mean is controlled by a solvable linear differential equation
\begin{equation}
\partial_t\overline{c} = \left(\mathcal{L}-2u_*\right)\overline{c}\;.
\label{eq:linfirstmoment}
\end{equation}
This equation is similar to the deterministic equation except for the negative last term that accounts for number fluctuations.

In summary, branching random walks occur whenever individuals proliferate, and are therefore a frequent element in both ecological and evolutionary models. A given ecological model often has an interpretation in an evolutionary context as well, if the dynamical variables are reinterpreted. It is usually difficult to analyze these models when populations are regulated in size, for instance by a logistic non-linearity. The algorithm described above provides a method to construct a \emph{solvable} model of constrained branching random walks, in which the population size is fluctuating but finite. The algorithm to find and analyze this model consists of first solving equation (\ref{eq:ustar}) for the weighting function $u_*(x)$. Secondly, one has to determine the mean population density $\overline{c}(x,t)$ by solving the linear equation (\ref{eq:linfirstmoment}), which depends on the previously determined weighting function $u_*(x)$.  It has been shown in \cite{hallatschek2011noisy} that this weighting function has a natural probabilistic interpretation: $u_*(x)$ is the \emph{fixation} probability that the descendants of an individual sampled from position $x$ will eventually take over the population. It has also been found in both Refs.~\cite{hallatschek2011noisy,good2012distribution} that the tuned model is useful to determine those properties, which are only weakly dependent on the form of the constraint.  In this way, it was possible to determine the speed of traveling waves in an primarily analytic way. In the following, we will use this approach of model tuning to derive the behavior of the death rate in our stochastic oasis model.

\subsection{Stochastic oasis model}
\label{sec:oasis-model}

When the Liouville operator $\mathcal{L}$ in (\ref{eq:linoperator}) corresponding to our oasis problem is used, the steady state equations for $u_*(x)$ and for $\overline c_{st}(x)$ take the form
\begin{eqnarray}
\label{eq:ustar-eqn-orig}  0 &=& D\partial_x^2u_*- v\partial_x u_*+ (\alpha \delta(x)- r)u_*-2 u_*^2\;, \\
\label{eq:cmean-eqn-orig}0 &=& D\partial_x^2 \overline{c} + v\partial_x  \overline{c}+ (\alpha \delta(x)- r) \overline{c}-2 u_*  \overline{c}\;.
\end{eqnarray}
To simplify the notation in the following, we introduce rescaled variables. We use $X\equiv x \alpha/D$ to measure length in units of $D/\alpha$. We also introduce new variables for the population density $C(X)\equiv c(X D/\alpha) \alpha$ and the weighting function $U_*(X)\equiv u_*(X D/\alpha) D/\alpha^2$. The equations of motion
\begin{eqnarray}
\label{eq:ustar-eqn}  0 &=& \partial_X^2 U_*- \nu\partial_X U_*+ \left[\delta(X)- \frac{\rho}4\right]U_*-2 U_*^2\;, \\
\label{eq:cmean-eqn}0 &=& \partial_X^2 \overline{C} + \nu\partial_X  \overline{C}+ \left[\delta(X)- \frac{\rho}4\right] \overline{C}-2 U_*  \overline{C}\;.
\end{eqnarray}
together with the constraint
\begin{equation}
  \label{eq:const}
  1=\int dX~U_*(X) C(X)
\end{equation}
then form a closed set of equations that only depends on two control parameters, the rescaled velocity $\nu\equiv v/\alpha$ and the rescaled death rate $\rho\equiv 4D r \alpha^{-2}$.~\footnote{The rescaled death rate was chosen such that, in the deterministic limit, we have  $\rho=1$ in the convection less case.
The rescaled velocity was chosen such that the deterministic delocalization threshold is at $\nu=1$.}

\subsubsection{\label{sec:withoutconvection}Stochastic oasis model without convection}

It turns out that the convection terms in (\ref{eq:ustar-eqn}) and (\ref{eq:cmean-eqn}) considerably complicate the analysis. Therefore, we first focus on the case without convection, $\nu=0$, and treat the windy situation in approximation later on, in section \ref{sec:withwind}.
  
The equation for $U_*$ with $\nu=0$ restricted to $X>0$ can be mapped onto the mechanical problem of a point particle in a potential. When we interpret $X$ as time and $U_*$ as the location of the particle, its dynamics follows Newton's equation
\begin{equation}
  \label{eq:Newton}
  \partial_X^2 U_*=-\partial_{U_*} V(U_*)\;,
\end{equation}
where the potential $V(U_*)$ is given by
\begin{equation}
  \label{eq:potential}
  V(U_*)=- 2 U_*^3/3-   \rho U_*^2 /8 \;.
\end{equation}
Using this analogy, we obtain the weighting function $U_*(x)$ through the integral
\begin{equation}
dX = \frac{dU_*}{\sqrt{2\left(E-V(U_*)\right)}} \;,
\label{eq:trajectory_differential}
\end{equation}
where $E$ represents the sum of kinetic and potential energy of the moving particle. The value of $E$ is fixed by the requirement that  the ``trajectory'' (here proportional to the density profile) approaches zero infinitely slowly in the limit $X\rightarrow\infty$.
Thus, the total energy must equal the potential energy at $U_*=0$. Hence, $E=V(0)=0$. The integral in (\ref{eq:trajectory_differential}) from $U_*(0)$ to $U_*(X)$ can then be performed analytically, which yields
\begin{equation}
\label{eq:u*-solution-general}
U_*(X) = \frac{3\rho}{16}\left[\sinh\left(\sqrt{\rho}X/4-\mbox{arcsch}\left(\sqrt{\frac{16U_*(0)}{3\rho}}\right)\right)\right]^{-2}\;.
\end{equation}
The value $U_*(0)$ has to be fixed by the boundary condition at $X=0$. This boundary condition is obtained from the full equation for $U_*$, equation (\ref{eq:ustar-eqn}), by integrating from $X=-\epsilon$ to $X=\epsilon\ll1$, 
\begin{equation}
  \label{eq:boundary-cond}
  \left[ U'_*(\epsilon)-U'_*(-\epsilon) \right]+U_*(0)=0 \;,
\end{equation}
which also holds for $\nu>0$. In the absence of the convection term, we expect $U_*$ to be an even function. Therefore, we must demand the boundary condition 
\begin{equation}
  \label{eq:bc_at_0}
  U'_*(0) = -U_*(0)/2 \;.
\end{equation}
This boundary condition is satisfied by our solution (\ref{eq:u*-solution-general}) when we set
\begin{equation}
  U_*(0) = \frac{3}{16}\left(1-\rho\right)\;.
\label{eq:offset}
\end{equation}
Due to the symmetry in $X$, the full solution for $U_*$ is given by
\begin{equation}
\label{eq:ustar-solution}
U_*(X) = \frac{3\rho}{16}\left[\sinh\left(\frac{\sqrt{\rho}}{4}\vert X\vert + \mbox{artanh}\left(\sqrt{\rho}\right) \right)\right]^{-2}\;.
\end{equation}
Note that the form of the equations  (\ref{eq:ustar-eqn},~\ref{eq:cmean-eqn}) for $U_*$and $\overline C$ implies that $U_*(x)$ is proportional to a stationary density profile,
\begin{equation}
\label{eq:predicted_profile}
\overline{C}_{st} \sim U_*\;,
\end{equation}
where the proportionality constant $\left(\int dX~U_*^2(X)\right)^{-1}$ is fixed by the constraint in (\ref{eq:const}).

\begin{figure}[!t]
\begin{center}
\includegraphics[width=10cm]{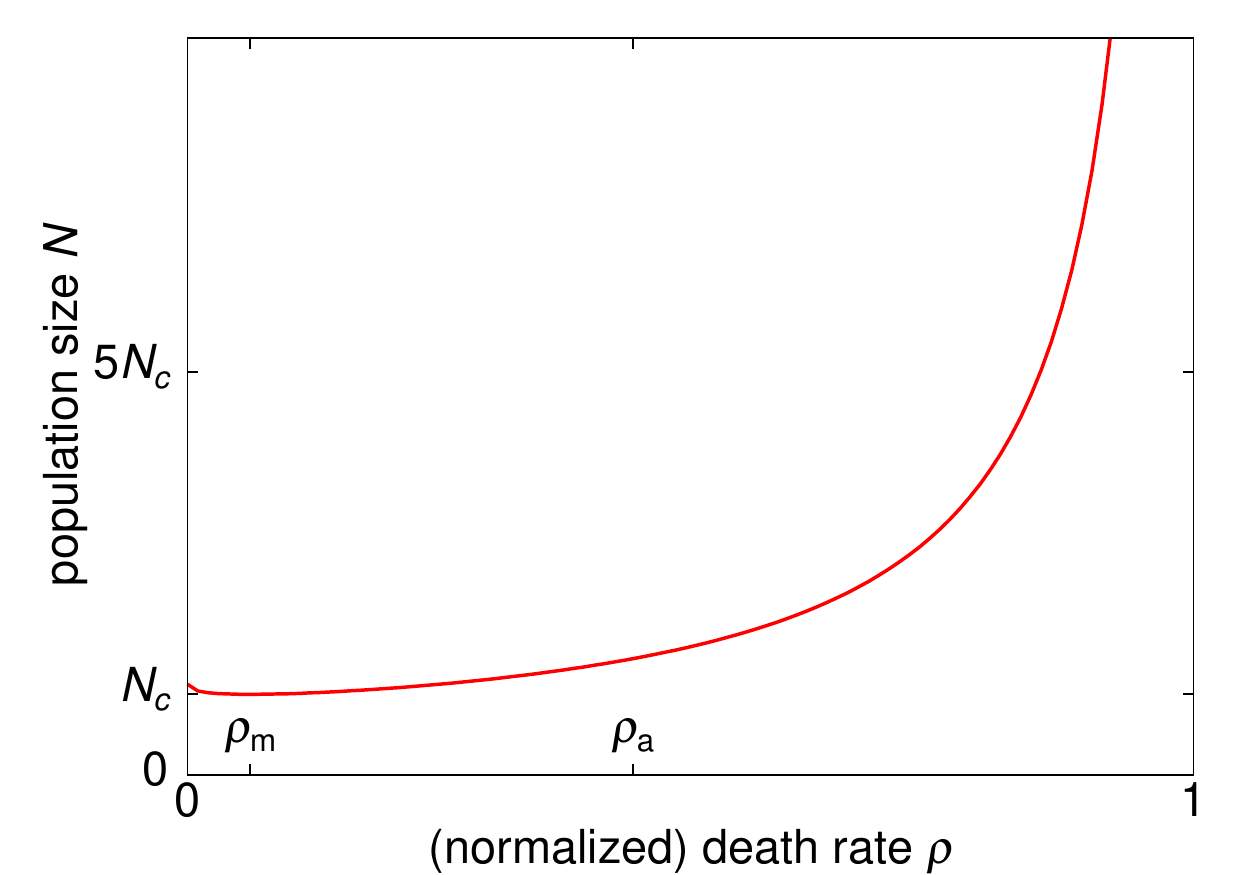}
\caption{\label{fig:meanpopsize}
\textbf{Mean population size $\overline{N}$ obtained for the analytical model with tuned constraint, in the absence of convection, $\nu=0$.}
The population size diverges at $\rho=1$, which corresponds to the deterministic limit. For $\rho<\rho_a\approx0.44$, the population profile $\overline c_{st}$ acquires a pronounced algebraic decay up to a length $\xi\sim \rho^{-1/2}$. This broad tail arises from large scale center-of-mass fluctuations. At $\rho=\rho_m=1/16$ the mean population size has a minimum. }
\end{center}
\end{figure}

By integrating over the density profile, $\overline N = (D/\alpha^2)\int dX~\overline C_{st}(X)$, we obtain a relation between the  mean population size  $\overline N$ and the rescaled death rate $\rho\equiv 4Dr\alpha^{-2}$,
\begin{equation}
\overline{N} = \frac{D}{\alpha^2}\frac{16}{1+\sqrt{\rho}-2\rho}\;.
\label{eq:meanpopsize}
\end{equation}
The predicted dependence of the mean population size on the death rate $\rho$ is depicted in Figure \ref{fig:meanpopsize}.  Notice that the (mean) population size $\overline{N}$ reaches a minimum at $\rho=\rho_m=1/16$. This minimum is a distinctive feature of the tuned model close to the delocalization transition. In this regime, the population frequently detaches from the oasis and takes large excursions away from the oasis until it is eventually pushed back to the oasis. During such excursions, the population explores the tail of the function $u_*$. To obey the constraint (\ref{eq:generalizedconstraint3}) for a low value of $u_*$, the population has to be rescaled to a large value, which leaves its footprint in an overall increased mean population size. Since this phenomenon cannot occur in a model with $u=$const., fixed population size models do not exhibit such a minimum in the population size.

\subsubsection{Deterministic and stochastic limits }
\label{sec:limiting-cases}
\begin{figure}[!tb]
\begin{center}
\includegraphics[width=10cm]{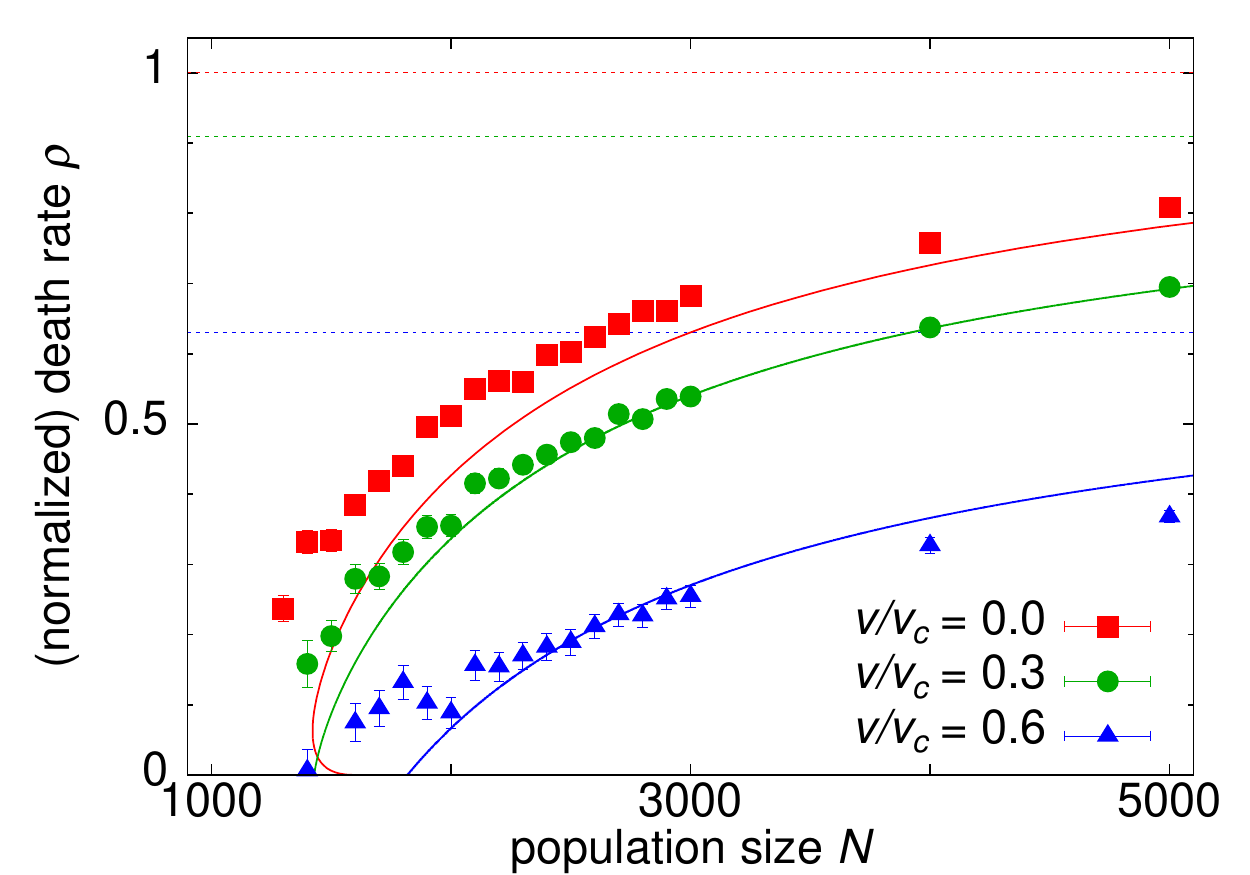}
\caption{\label{fig:Nvsrho}\textbf{Relation between death rate $\rho$ and population size $N$ for different convection velocities $v$.} Symbols represent measured death rates in the fixed population size model for different population sizes and convection velocities (see legend). Dashed horizontal lines are the deterministic approximations (\ref{eq:rho-deterministic}), which strongly deviate from the simulations of finite populations.
The solid red line is the predicted (mean) population size $\overline N$ for a given death rate $\rho$, c.f. (\ref{eq:meanpopsize}), derived from the solvable model with tuned constrained in the convection-less case. Notice that these predictions also capture the quantitative trend of the simulation data for fixed population sizes (red symbols).
The green and blue lines are approximations to the tuned model with convection, c.f. (\ref{eq:sva-r}). Simulation parameters are $a=0.1$, $\epsilon=0.2$.}
\end{center}
\end{figure}

Our solutions for $U_*$ and $\overline C_{st}$ are valid as long as
\begin{equation}
0< \rho\equiv \frac{4Dr}{\alpha^2} < 1\;.
\label{eq:conditionsonr0}
\end{equation}
The bounds in (\ref{eq:conditionsonr0}) can be readily interpreted. For  $\rho\to 1 $, we recover the exponential decay known from the deterministic limit (\ref{eq:ss-profile-deterministic}),
\begin{equation}
\overline{C}_{st}(X) \sim \exp\left(-\vert X\vert/2\right)\;.
\end{equation}
The total population size scales as $\overline{N}\sim32D/(3\alpha^2\delta \rho)$ for $\delta \rho\equiv 1-\rho\ll1$.  Since relative fluctuations in population size become small for large populations, we expect this leading order result  to be independent of the form of the constraint. Indeed, the measured growth rates in the fixed $N$ model for $\nu=0$ agree very well with our prediction, as can be seen from Figure~\ref{fig:Nvsrho}.

\begin{figure}[!th]
\begin{center}
\includegraphics[width=10cm]{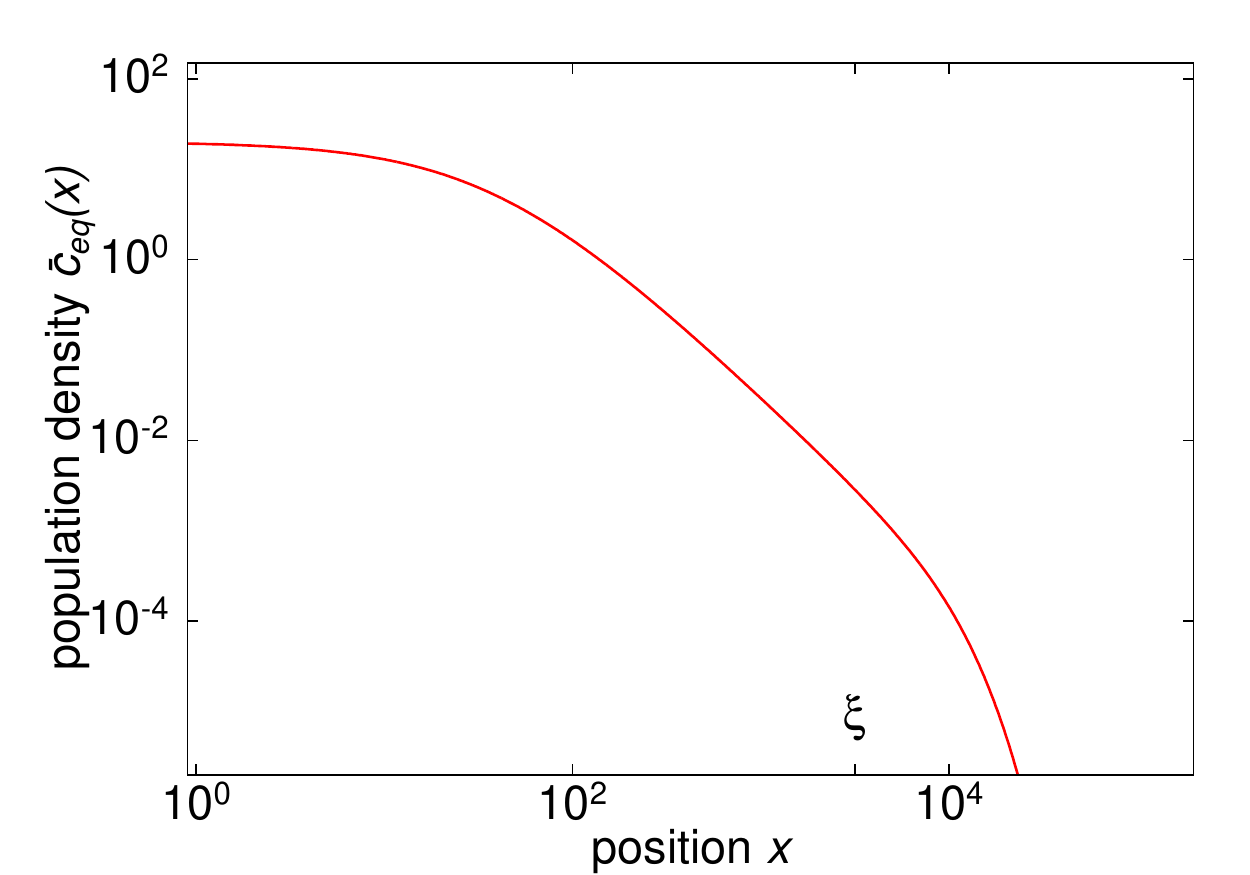}
\caption{\label{fig:crossover}\textbf{Density profile $\overline{c}_{st}(x)$ close to the stochastic threshold, $\rho\ll1$.} The plot depicts the density profile in (\ref{eq:predicted_profile},~\ref{eq:ustar-solution}) predicted from the solvable model with tuned constraint. We used the parameters $\alpha=0.1, r=10^{-7}, D=1$, such that the rescaled death rate is much smaller than one, $\rho\equiv 4Dr\alpha^{-2}=4\times 10^{-5}$.
Notice the intermediate algebraic decay (with exponent $-2$) up to the correlation  length $\xi\sim 2\sqrt{\frac{D}{r}}$. }
\end{center}
\end{figure}

In the case of very small death rates, $\rho\ll1$, we obtain an algebraic decay with distance,
\begin{equation}
\overline{C}_{st}(X) \sim \frac{1}{2}\left(1+\frac{|X|}4\right)^{-2}\;,
\label{eq:algebraic-decay}
\end{equation}
This algebraic decay indicates that the center-of-mass of the population fluctuates strongly around the center of the oasis. 
Figure \ref{fig:crossover} depicts the density profile for small but finite $\rho$, for which the decay is algebraic (as in (\ref{eq:algebraic-decay})) up to a characteristic crossover scale (in original units)
\begin{equation}
\xi \sim 2\sqrt{\frac{D}{r}}\;.
\label{eq:crossover_scale}
\end{equation}
For $x \gg\xi$, the decay is exponential. We note that to clearly see the algebraic tail, the rescaled growth rate has to be finely tuned to a very small value. Yet, an algebraic decay of correlations and a diverging correlation length, is reminiscent of a second order phase transition. This raises the question of whether the underlying physics is generic for a wide range of systems. 

Some more insight can be gained with the help of a simple scaling argument for the extinction dynamics close to the transition. For very small death rates, assume that the oasis sheds off subpopulations into the surrounding environment. Since this desert is only mildly deleterious, these trailing populations will survive up to a maximal time of roughly $1/r$.
Within this time, the subpopulations diffuse up to a length scale $\sqrt{D/r}$. Incidentally, this is precisely the correlation length, defined in (\ref{eq:crossover_scale}), where the density profile crosses over from algebraic to a much faster exponential decay. The algebraic decay of the density profile within the correlation length can be explained by involving the statistics of branching random walks.
Suppose, the populations are shed off from the oasis at a small, but constant rate $k_{off}$. Each of these populations follows a critical branching process as long as the population is smaller then $1/r$.
(For larger populations, selection dominates and leads to a continual decrease in population size at rate $r$.) First, the probability for a small population to reach a size larger than $N$ individuals through a critical branching process is proportional to $1/N$. The time to reach the population size $N$, conditional on survival, is roughly given by $N$ generations. During this time, the position $X_{com}$ of the center-of-mass will have diffused by a length scale $\sqrt{D N}$.
Combining these scaling arguments, we determine the probability that a sub-population reaches further than the distance $x$ as
\begin{equation}
  \label{eq:CM_diffusion}
  Pr(X_{com}>x)\sim Pr(T>x^2/D) \sim Pr(N>x^2/D) \sim \frac{D}{x^2}\;.
\end{equation}
Now, the mean population density at $x$ will be given by the probability density to reach $x$, $Pr(X_{com}=x)\sim D x^{-3}$, times the typical density of a population that reaches $x$. Such a population has size $N_x\sim x^2/D$ and a lateral spread of $\sqrt{D N_x}$ (acquired during a coalescence time of $N$ generations). Hence, the density of population that reaches $x$ scales as $x/D$, and we obtain
\begin{equation}
  \label{eq:density}
  \overline c(x)\sim Pr(X_{com}=x) x/D\sim x^{-2}\;.
\end{equation}
The pre-factor clearly depends on the rate at which populations are shed off from the oasis. 

The above scaling picture suggests, that the exponent $-2$ is generic for a \emph{critical} branching random walk with a boundary condition of a finite (source) population on one end. In this context, critical means that birth and death rate are equal. If the landscape is only \emph{nearly} critical and characterized by a weak selective imbalance $r$, then we expect to still see the power law up to the cutoff $\sqrt{D/r}$.

\subsection{\label{sec:withwind} Stochastic oasis model with convection}
A finite wind speed significantly complicates the mathematical analysis, because it results in a non-hermitian  Liouville operator, defined in (\ref{eq:linoperator}). As long as the population remains localized, it is however possible to return to a hermitian problem by the non-linear variable transformation,
\begin{eqnarray}
  \label{eq:psi}
U_*(X)& =& e^{\frac{\nu X}{2}}\Psi(X)\;,\\
  \label{eq:phi}
\overline{C}(X) &= &e^{-\frac{\nu X}{2}}\Phi(X)  \;.
\end{eqnarray}
In terms of the new functions $\Psi(X)$ and $\Phi(X)$, the stationary equations of motion read
\begin{eqnarray}
\label{eq:transformed_psi} 0 &=& \partial_X^2\Psi(X) +\left[\delta(X) - \frac{\rho}{4}-\frac{\nu^2}{4}\right]\Psi(x)-2 e^{\frac{\nu X}{2}}\Psi^2(X) \;,\\
\label{eq:transformed_phi}0 &=& \partial_X^2\Phi(X)+\left[\delta(X) - \frac{\rho}{4}-\frac{\nu^2}{4}\right]\Phi(x) -2 e^{\frac{\nu X}{2}}\Psi(X)\Phi(X) \;.
\end{eqnarray}
These equations cannot be solved in closed form, but a simple approximation is possible for small convection velocities where the exponential factors in (\ref{eq:transformed_psi},\ref{eq:transformed_phi}) are close to $1$. This approximation is justified if the exponent is smaller than $1$ on the considered length scale. If we consider the localization length of the deterministic problem (c.f. (\ref{eq:correlation-scale})) as the relevant scale, this implies that the convection speed should be much less than the deterministic threshold, $\nu\ll1$.

After replacing the exponential factors in (\ref{eq:transformed_psi}, \ref{eq:transformed_phi}) by $1$, the equations of motion become identical to the convection free case with the replacement $\rho\mapsto \rho+\nu^2$. Accordingly, the density profiles then obey
\begin{equation}
\overline{C}_{st}(X;\rho,\nu) \approx\overline{C}_{st}\left(X;\rho+\nu^2,0\right)e^{-\frac{\nu X}{2}} \;, \qquad \nu\ll 1  \;.
\label{eq:sva}
\end{equation}
The resulting predictions for the population densities and growth rates have been used in Figure~\ref{fig:densshapes} to compare with our fixed $N$ simulations. The agreement is not perfect, due to the combined error generated by the model difference and our small velocity approximation, but reproduces well the qualitative behavior of the simulations.

\subsubsection{\label{sec:appr-death-rate}Approximation for the death rate}
Finally, we use the tuned model to derive a rather simple prediction for the death rate $\rho$ as a function of population size and convection speed. This approximation will help us better understand the delocalization mechanism in the presence of number fluctuations. 

Integrating the governing equation~(\ref{eq:cmean-eqn-orig}) for the mean density profile $\overline{c}(x)$ over the space coordinate $x$ yields
\begin{equation}
  \label{eq:r-approx1}
  0=-r \int dx~ \overline{c}(x)+\alpha\overline{c}(0)-2\int dx~ u_*(x)\overline{c}(x)\;. 
\end{equation}
Since the first term is proportional to the total population size $\overline N$, and the last term is fixed by the constraint (\ref{eq:generalizedconstraint3}), we obtain in general
\begin{equation}
  \label{eq:r-approx2}
  r=\frac{\alpha \overline{c}(0)}{\overline N}- \frac{2}{\overline N}\;.
\end{equation}
This result is identical to the mean field prediction except for the negative last term.  Hence, the death rate is lower than the mean-field prediction by the term $2/\overline N$. As a consequence, smaller convection speeds than in the mean-field approximation are required for the death rates to cross $0$, which signals the delocalization of the population. 

For using equation (\ref{eq:r-approx2}) predict the death rates for a given parameter set, we still need to find the occupancy $\overline c(0)$ at the oasis. This task requires the non-trivial solution of the set of equations in (\ref{eq:transformed_psi}). By employing the small-velocity approximation (\ref{eq:sva}) to express $\overline c(0)$ in (\ref{eq:r-approx2}), we obtain
\begin{equation}
r\approx\left(\frac{\alpha}{8\sqrt{D}}+\sqrt{\left(\frac{3\alpha}{8\sqrt{D}}\right)^2-\frac{2}{\overline N}}\right)^2-\frac{v^2}{4D}\;,
\label{eq:sva-r}
\end{equation}
This approximation (\ref{eq:sva-r}) is used in Figures \ref{fig:deathrates}, \ref{fig:Nvsrho} to explain fixed $N$ simulation data, and ultimately also to predict the fuzzy phase boundary in Figure \ref{fig:locstates}, as implicit condition when the death rates reaches zero.

\section{Discussion}
\label{sec:Discussion}
We have studied localization of a finite population in an ecological context. Within our model, individuals enjoy an effective growth rate benefit in one site of a linear chain of subpopulations. We asked the question under which conditions this preferred growth spot, which has been called ``oasis'', might be able to pin the population. We found that, under a model of globally regulated total population size, escape from the oasis inevitably occurs due to random number fluctuations.
The escape time however sensitively depends on both the population size and the convection speed. The delocalization dynamics may be summarized in the phase diagram of Figure~\ref{fig:locstates}. When the convection speed is small enough and the population large enough, escape is extremely unlikely and the population can be considered as effectively localized. Due to the exponential dependence of escape times on $N$ and $v$, the cross-over from long to short delocalization times is very sharp in the $N$--$v$ plane.
The measured density profiles converge to the exponential decay for large population sizes, as predicted from deterministic mean field theories \cite{dahmen2000life,desai2005quasispecies}. The agreement with the deterministic theory generally becomes poorer as escape events become more frequent, either through an increase in convection velocity or through a reduction in population size. Deviations are pronounced in particular in the expected death rate that has to be used to control the population.
In the absence of convection, this death rate is found to continually decrease as the population size is lowered. This is in contrast to the population size independent death rate predicted by the mean-field theory.

To analyze these findings, we have solved a variant of the model with a tuned population size constraint. In this model, a weighted sum over the occupancies of all demes is held strictly constant. The weighting function is chosen such that the first moment equation is linear. This model has the advantage of being analytically tractable. We found that the tuned model exhibited a sharp delocalization transition for all population sizes.
The phase transition line in the $N$--$v$ plane could be determined in a simple analytical approximation, which is indicated by the dashed red line in the phase diagram Figure~\ref{fig:locstates}. Interestingly, we found that our tuned model exhibits a continuous delocalization transition driven solely by number fluctuations, even in the absence of convection. Close to the transition, the density profile exhibits a power law tail with exponent $-2$ up to a correlation length that diverges at the transition. 

It has to be remarked that the reason for the sharp phase transition in the tuned model is ultimately due to the inhomogeneity of the weighting function. Occupancy of the oasis is weighted more strongly than occupancy of lattice sites away from the oasis. This is in contrast to the fixed population size constraint, which has a spatially homogeneous weighting function: The occupancy of all demes are summed up, independent on their identity, and compared with the pre-fixed value.
Therefore, the stochastic escape from the oasis is a characteristic effect of this degenerate weighting function. It remains to be seen, how one could explain analytically the exponential dependence of the typical escape time on population size and convection speed. Nevertheless, our simulations show that the narrow cross-over region from small to large escape times is located close to the sharp phase transition line of our tuned model in Figure~\ref{fig:locstates}.
Furthermore, our tuned model predicts reasonably well the behavior of the death rate required to keep the population size finite. Compared with the mean-field approximation, the death rate is lowered by a value of roughly $2/N$, in fair agreement with fixed $N$ simulations, c.f. Figure~\ref{fig:deathrates}.

Our results also have a natural interpreted in an evolutionary context. Consider a population evolving in a fitness landscape characterized by an isolated fitness peak. Mutations cause random movements within this fitness landscape, typically biased in certain directions, and reproduction is often approximated by a branching process subject to the constraint of a fixed population size.
The resulting quasi-species model of a finite population is a higher dimensional version of what we have considered in this paper, but otherwise analogous. We therefore expect our results could be useful for advancing the quasi-species theory, whose stochastic version has so far remained largely unexplored.

\section*{Acknowledgments}
\addcontentsline{toc}{section}{Acknowledgments}
We would like to thank Jens Nullmeier and Paulo Pinto for many valuable discussions. We also acknowledge the anonymous referees for helpful comments on the manuscript. This work was supported by the Max Planck Society.

\appendix

\renewcommand{\thesection}{\Alph{section}}

\section{Different choices for the definition ``population on the oasis''}
\label{apdx:condition}
\begin{figure}[!ht]
\begin{center}
\includegraphics[width=8cm]{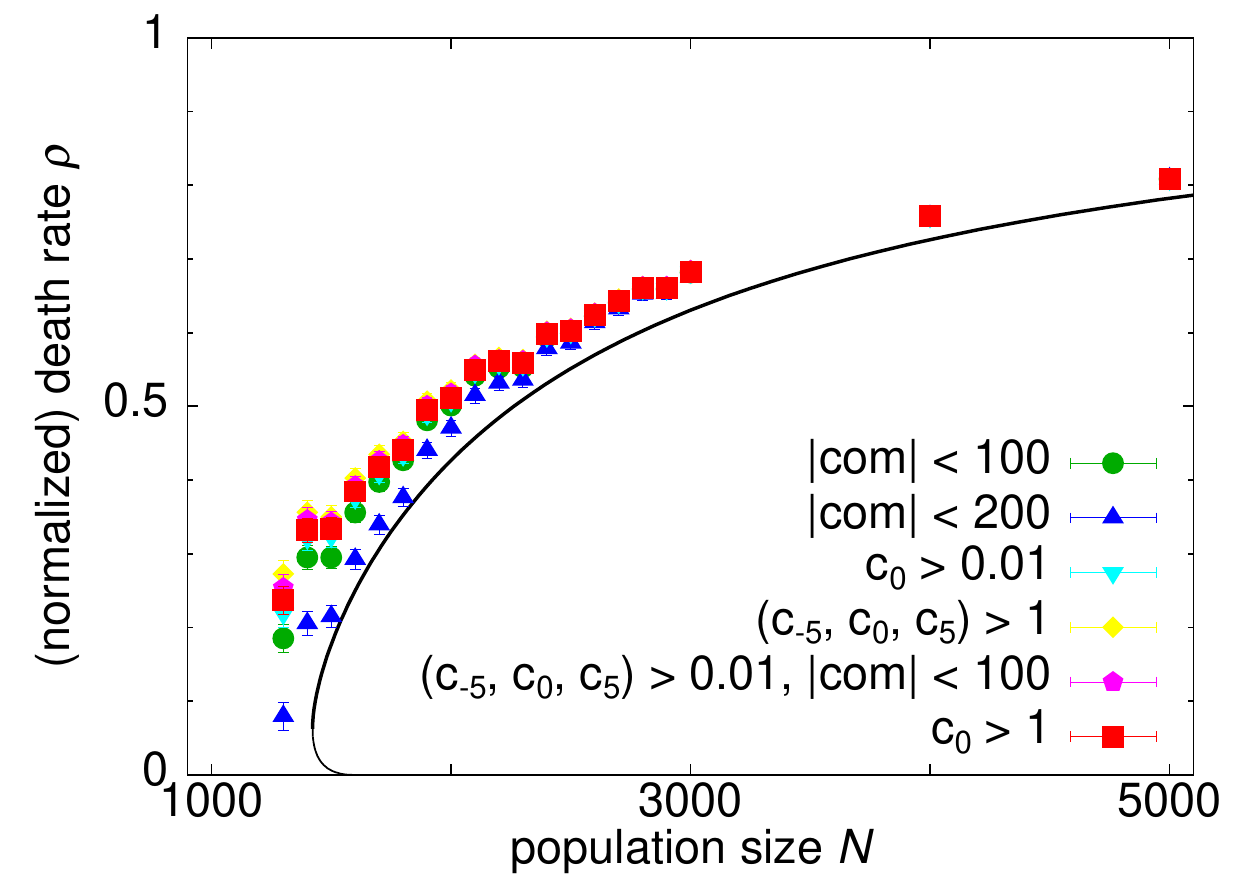}
\caption{\label{fig:onoff}\textbf{Different criteria for localization yield qualitatively similar results.} In the main text, the population was defined to be localized at the oasis if at least one individual was present at the oasis, $c_0>1$. Different criteria are of course possible. Here, we compare the following conditions: ``$\vert com\vert< X$'': the center-of-mass (com) is within a distance $X$ of the oasis. ``$(c_{-5}, c_0, c_{+5})>x$'': the  occupancies in demes $-5$, $0$ (oasis) and $5$ exceeds $x$ simultaneously. We also checked a combination of the last two conditions. Overall we find that our choice, $c_0>1$, is  in good agreement with the others, and the predictions from the tuned constraint model (black line).}
\end{center}
\end{figure}

In our simulations, detailed in Section \ref{sec:simulation-model}, the population always delocalizes from the growth spot if they are run long enough. Our analytically tractable model, on the other hand, describes populations that never fully detach from the oasis, provided the model parameter are within the regime in which the equations of motion, (\ref{eq:ustar-eqn},~\ref{eq:cmean-eqn-orig}), have non-trivial solutions. To compare both approaches, we needed to average observables only over those configurations in the simulation, which can reasonably be called ``localized''.  In Figure \ref{fig:onoff}, we present the death rate for several alternative choices for the definition of localized states.  We termed states localized if the center-of-mass (``com'') of the population was within a certain distance to the oasis and/or if the site of the oasis has an occupancy $c_0$ larger than a certain threshold value. The tested definitions gave very similar results. For definiteness, we chose $c_0>1$ for presenting averaged simulation data in the main text.  Note that, with this definition, the death rate averaged over localized states cannot entirely decay to zero at the delocalization transition. This together with the model differences are reasons for the discrepancy between fixed $N$ simulations and our analysis of the tuned model, as seen in Figures \ref{fig:deathrates}, \ref{fig:Nvsrho} and \ref{fig:onoff}.

\section{Relaxation spectrum in the case without convection}
\label{apdx:spectrum}
In the main text, we have analyzed the steady state of the analytically tractable tuned model. Here, we discuss the relaxation towards the steady state for the case without convection. The relaxation dynamics is governed by the eigenmodes and eigenvalues of the linear differential equation for $\overline{C}$~\cite{dahmen1999population,dahmen2000life,shnerb2001extinction}. The following analysis shows that only a single bound state exists, together with a continuum of extended, delocalized states. These extended modes decay exponentially, with a rate faster than $\rho/4$ in (rescaled) time.

\begin{figure}[!ht]
\begin{center}
\subfloat{\label{fig:spectrum:eigenvalues}\includegraphics[width=7.8cm]{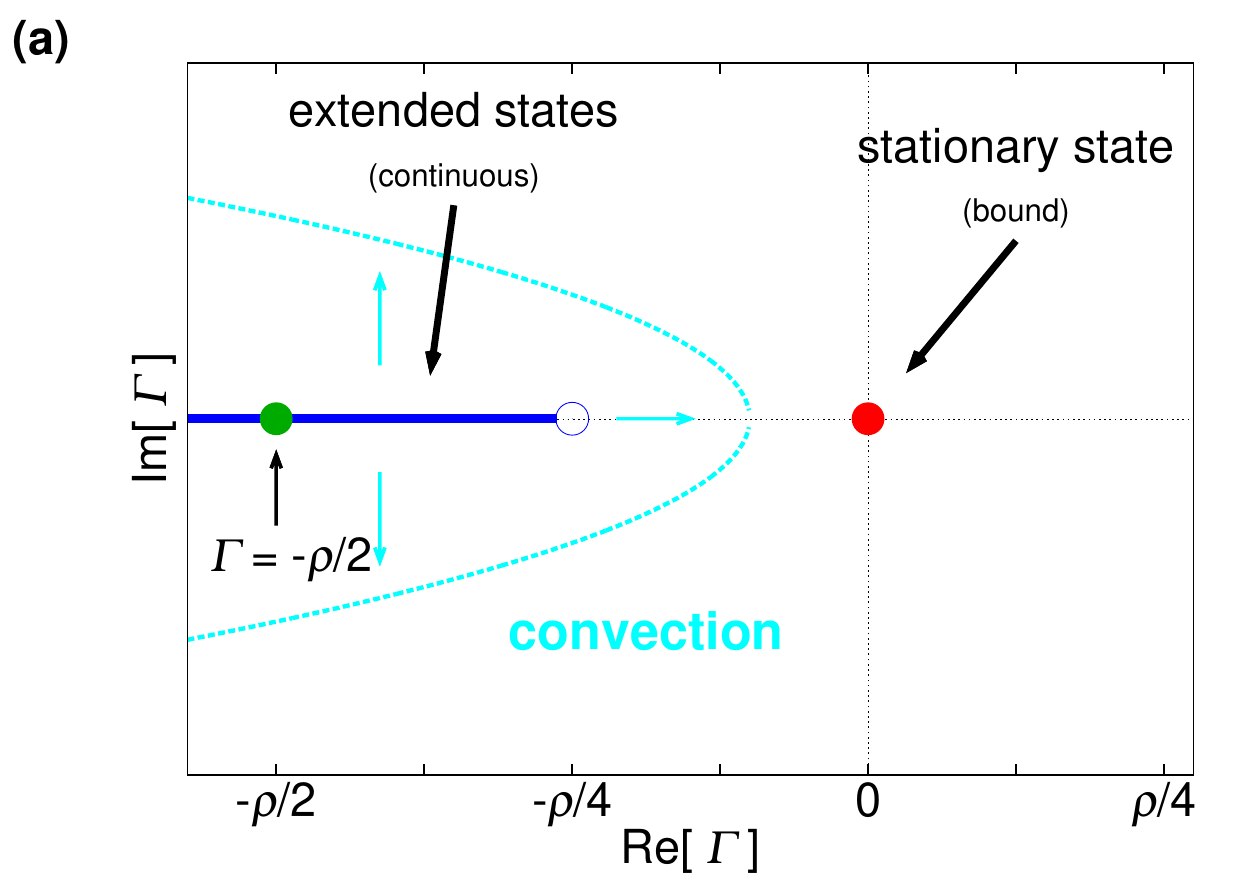}}
\subfloat{\label{fig:spectrum:eigenstates}\includegraphics[width=7.8cm]{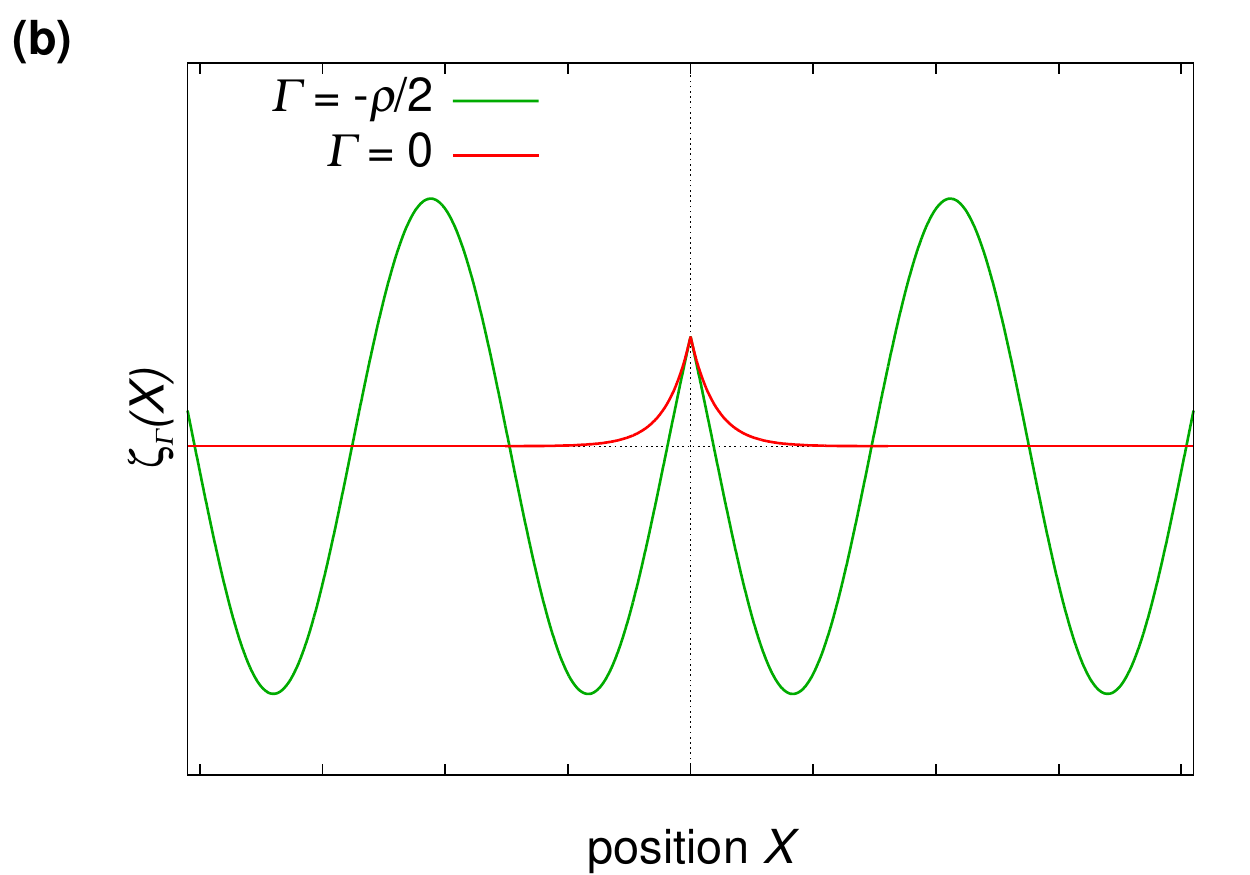}}
\caption{\label{fig:spectrum}\textbf{Spectral properties of the oasis
model.} (a) Without convection, $\nu=0$, the spectrum can be obtained
analytically and consists of the single bound state at $\Gamma=0$ and a
continuous line of eigenvalues, the open interval $\Gamma\in
\left(-\infty,-\rho/4\right)$, for the extended delocalized states. In
analogy with \cite{dahmen2000life} we expect the convection to render
the eigenvalues complex (in pairs). The real part of these eigenvalues
is approaching the bound state with increasing convection $\nu$ until
the first extended state reaches the bound state at the the
delocalization velocity $\nu_c$. (b) Bound state ($\Gamma=0$) and one of
the delocalized states ($\Gamma=-\rho/2$). Both states are normalized to
fulfill $\zeta_0(0)=\zeta_{-\rho/2}(0)$.}
\end{center}
\end{figure}

The dynamics of the mean density is described by (c.f.
(\ref{eq:linfirstmoment}) and (\ref{eq:linoperator}))
\begin{eqnarray}
\partial_t \overline C&=&\left(\mathcal{L}-2U_*\right)\overline{C}\;, \\
\mathcal{L} &=& \partial_X^2-\nu\partial_X +\delta(X)-\rho/4\;.\label{eq:oper}
\end{eqnarray}
The mean density $\overline{C}$ can be decomposed into a basis
consisting of (right) eigenfunctions of the linear operator
$\left(\mathcal{L}-2U_*\right)$. Without convection this operator is
hermitian, rendering all eigenvalues and eigenfunctions real valued.
Here, we only treat the convection-less case, $\nu=0$. The
eigenfunctions $\left\lbrace \zeta_\Gamma(X)\right\rbrace$ can then be
chosen as an orthogonal basis, satisfying
\begin{eqnarray}
\label{eq:eigenstates} \left(\mathcal{L}-2U_*\right)\zeta_\Gamma =
\Gamma\zeta_\Gamma\;, \\
\label{eq:eigenstates-normalization}\int dX~\zeta_
\Gamma(X)\zeta_{\Gamma'}(X) = 0~~~\mbox{if }\Gamma\neq\Gamma'\;.
\end{eqnarray}
In addition to (\ref{eq:eigenstates-normalization}) we can demand the normalization of $\zeta_0$, $\int dX \zeta_0(X)^2 = 1$, which is used later in (\ref{eq:spectrum:constr:decomp}).

By the spectral theorem we can write
\begin{equation}
\overline C(X,t) = \sum\limits_\Gamma \kappa_0(\Gamma)\rme^{\Gamma
t}\zeta_\Gamma(X)\;,
\label{eq:spectrum:relaxationdynamics}
\end{equation}
characterizing the relaxation dynamics of perturbations in the mean
density. In (\ref{eq:spectrum:relaxationdynamics}) the coefficients
$\left\lbrace \kappa_0(\Gamma)\right\rbrace$ are the projections of the
mean density $\overline C$ at initial time $0$ onto the eigenstates
$\zeta_\Gamma$,
\begin{equation}
\kappa_0(\Gamma) = \int dX~\overline C(X,0)\zeta_\Gamma(X)\;.
\end{equation}
Note that our constraint fixes the coefficient of the stationary
eigenfunction with eigenvalue zero,
\begin{eqnarray}
\label{eq:spectrum:constr:original}1 &=& \int dX~U_*(X)\overline C(X,t)
\\
\label{eq:spectrum:constr:decomp}&=& \beta\sum\limits_\Gamma
\kappa_0(\Gamma)\rme^{\Gamma t}\int dX~\zeta_0(X)\zeta_\Gamma(X) \\
\label{eq:spectrum:constr:final}&=& \beta \kappa_0(0)\;.
\end{eqnarray}
In going from (\ref{eq:spectrum:constr:original}) to
(\ref{eq:spectrum:constr:decomp}) we used the fact that $U_*$ is
proportional to $\zeta_0$ with a proportionality constant $\beta$.

Inserting the operator in (\ref{eq:oper}) into (\ref{eq:eigenstates}) leads to 
\begin{eqnarray}
\nonumber \left[\partial_X^2-\frac{3\rho}{8}\left(\sinh\left(\frac{\sqrt
\rho}{4}\vert X\vert+\mbox{artanh}(\sqrt
\rho)\right)\right)^{-2}\right]\zeta_\Gamma(X) \\
=\left(\Gamma+\frac{\rho}{4}\right)\zeta_\Gamma(X),
\end{eqnarray}
for $X>0$ and $\nu=0$.

The eigenfunction $\zeta_\Gamma$ is a linear superposition, $\zeta_
\Gamma(X) \equiv \zeta_\Gamma^{(+)}(X) + K\zeta_\Gamma^{(-)}(X)$, of two
functions $\zeta_\Gamma^{(+)}$ and $\zeta_\Gamma^{(-)}$, given by
\begin{eqnarray}
\nonumber \zeta^{(+)}_\Gamma(X) &\propto& 
\label{eq:spectrum:zetap}\left[2+\frac{16\Gamma}{3\rho} + \left( \sinh
\left(Y(X)\right)\right)^{-2} \right]\cosh\left(2\sqrt{\frac{4
\Gamma}{\rho}+1}~Y(X)\right) \\
& &  + 2\sqrt{\frac{4\Gamma}{\rho}+1} \coth\left(Y(X)\right)\sinh\left(2
\sqrt{\frac{4\Gamma}{\rho}+1}~Y(X)\right)\;, \\
\nonumber \zeta^{(-)}_\Gamma(X) &\propto& 
\label{eq:spectrum:zetam}\left[2+\frac{16\Gamma}{3\rho} + \left( \sinh
\left(Y(X)\right)\right)^{-2} \right]\sinh\left(2\sqrt{\frac{4
\Gamma}{\rho}+1}~Y(X)\right) \\
& &  + 2\sqrt{\frac{4\Gamma}{\rho}+1} \coth\left(Y(X)\right)\cosh\left(2
\sqrt{\frac{4\Gamma}{\rho}+1}~Y(X)\right)\;.
\end{eqnarray}
Here, we have introduced the abbreviation $Y(X) \equiv \frac{\sqrt
\rho}{4} \left\vert X\right\vert+\mbox{artanh}(\sqrt\rho)$. The parameter $K$ is fixed by the
two boundary conditions
\begin{eqnarray}
\label{eq:eigenstates-bc1}\lim\limits_{X\rightarrow\pm\infty}\zeta_
\Gamma(X) < \infty\;, \\
\label{eq:eigenstates-bc2}\zeta'_\Gamma(0)=-\zeta_\Gamma(0)/2\;.
\end{eqnarray}

Different cases for $\Gamma$ have to be distinguished now. $\Gamma \geq
-\rho/4$ and $\Gamma<-\rho/4$ render the term $\sqrt{\frac{4
\Gamma}{\rho}+1}$ in the eigenfunctions either real or purely imaginary,
respectively. First, we treat the case $\Gamma \geq -\rho/4$. The
boundary condition (\ref{eq:eigenstates-bc1}) is fulfilled only for
$K=-1$. For the second boundary condition (\ref{eq:eigenstates-bc2}) we
obtain $\Gamma=0$ as a necessary requirement, leading to $\zeta_0(X)
\propto \left[\sinh(Y(X))\right]^{-2}$. This is the (already known)
stationary state, $\overline C_{st}(X)\propto\zeta_0(X)$. For $\Gamma < -\rho/4$,
the arguments of the hyperbolic functions are (purely) imaginary. Using
the identity between hyperbolic functions with complex arguments and
trigonometric functions, we obtain oscillating (but bound) functions
$\zeta_\Gamma^{(\pm)}$, which automatically comply with the first
boundary condition. Straightforward, but tedious, calculations yield the
coefficient $K$, which depends only on $\Gamma$ and $\rho$.

These calculations imply that the spectrum of the linear operator
$\left(\mathcal{L}-2U_*\right)$ is given by
\begin{equation}
\Gamma \in \left\lbrace0\right\rbrace \cup \left(-\rho/4,-\infty
\right)\;,
\end{equation}
as illustrated in Figure \ref{fig:spectrum:eigenvalues}. $\Gamma=0$ is
the single bound state, whereas we obtain a continuum of extended,
oscillating functions, which correspond to the delocalized states.

Analogous to previous results on the deterministic case
\cite{dahmen2000life}, we expect that the real part of the eigenvalue of
the extended states increases with increasing convection speed. Hence,
perturbations relax more slowly than in the convection-less case. At the
critical convection speed the relaxation time diverges.

\section*{References}
\addcontentsline{toc}{section}{References}

\bibliographystyle{plain}

\end{document}